\documentclass{aa}  
\usepackage[utf8]{inputenc}

\usepackage{graphicx}

\usepackage{txfonts}

\usepackage{natbib}
\usepackage{amsmath}
\usepackage{multirow}
\usepackage{booktabs}
\usepackage{float}
\usepackage{hyperref}
\usepackage{graphics} 
\usepackage{subfig} 
\usepackage{xcolor}
\usepackage{enumitem}
\usepackage{amsmath}
\usepackage{bm}
\usepackage{mathtools}
\usepackage{soul}
\begin{document}

   \title{Planet Earth in reflected and polarized light}

   \subtitle{I. 3D radiative transfer simulations of realistic surface-atmosphere systems}

   \author{Giulia Roccetti
          \inst{1,2}
          \and
          Claudia Emde
          \inst{2,3}
          \and
          Michael F. Sterzik
          \inst{1}
          \and
          Mihail Manev
          \inst{2}
          \and
          Julia V. Seidel
          \inst{4,5}
          \and
          Stefano Bagnulo
          \inst{6}
          }

   \institute{European Southern Observatory, Karl-Schwarzschild-Straße 2, 85748, Garching near Munich, Germany\\
    \email{giulia.roccetti@eso.org}
    \and
    Meteorologisches Institut, Ludwig-Maximilians-Universität München, Munich, Germany
    \and
    Rayference, Rue d’Alost 7, 1000, Bruxelles, Belgium
    \and
    European Southern Observatory, Santiago, Chile
    \and
    Laboratoire Lagrange, Observatoire de la Côte d’Azur, CNRS, Universit\'e Côte d’Azur, Nice, France
    \and
    Armagh Observatory and Planetarium, College Hill, Armagh BT61 9DG, Northern Ireland, UK
    }

   \date{Received xx; accepted xx}

 
  \abstract{Future ground- and space-based telescopes will enable the characterization of rocky exoplanets in reflected light, allowing for the observation of their albedo, which depends on surface, cloud, and atmospheric properties. Identifying key atmospheric, cloud and surface features is essential for assessing the potential habitability of these exoplanets. We present reference spectra and phase curves for a spatially unresolved Earth-like exoplanet in reflected and polarized light, highlighting how wavelength- and phase-angle-dependent reflectance reveal key planetary properties. Performing simulations with the 3D Monte Carlo radiative transfer code MYSTIC, we improve surface and cloud modeling by introducing validated, wavelength-dependent albedo maps of Earth's seasonal and spectral features, as well as a novel treatment of sub-grid cloud variability and inhomogeneities based on reanalysis data from ERA5. Our models incorporate high-resolution 3D cloud structures, demonstrating that sub-grid cloud variability significantly affects both intensity and polarization. It reduces total reflectance and increases phase curve variability, especially at large phase angles where ocean glint dominates. Additionally, we show that neglecting realistic wavelength-dependent albedo maps leads to a significant overestimation of the vegetation red edge feature in reflected light spectra. Comparing an Ocean planet to an Earth-like planet with seasonal cloud variability, we find that polarization is far more sensitive than intensity alone to distinguish both scenarios. Moreover, polarization captures richer information on surface properties, making it a critical tool for breaking degeneracies in retrieval frameworks. We present detailed model simulations that provide a ground-truth reference for observing Earth as an exoplanet and serve as critical benchmarks for developing observational strategies and retrieval frameworks for future telescopes targeting small, rocky exoplanets. Furthermore, this study informs model requirements and establishes a framework to optimize strategies for characterizing rocky exoplanets, emphasizing the pivotal role of polarization in breaking retrieval degeneracies across different models.}


   \maketitle
%

\section{Introduction}
\label{sec:intro}

The next generation of ground-based and space-based observatories will attempt to characterize rocky exoplanets in the habitable zones \citep{kasting1993} of nearby stars. Reflected light observations offer a powerful method for studying these planets by analyzing starlight that is either reflected off their surfaces or scattered in their atmospheres. Although current reflected light measurements can only provide upper limits \citep{charbonneau1999, hoeijmakers2018, spring2022}, future instruments such as the ANDES \citep{palle2023} and PCS \citep{kasper2021} on the Extremely Large Telescope, along with the proposed Habitable Worlds Observatory (HWO) \citep{US_decadal_survey}, will enable the detection of reflected light from rocky exoplanets.\\
\noindent Reflected light is sensitive to planetary albedo, which is influenced by surface features such as ocean glint \citep{robinson2011, Livengood2011, Robinson2014, lustig-yaeger2018, emde2017, ryan2022, trees2022, vaughan2023} and the vegetation red edge (VRE) \citep{Arnold2002, woolf2002, seager2005, tinetti2006b, fujii2011, kawahara&fujii2010, Fei2021} as well as atmospheric properties like clouds \citep{kaltenegger2007, kitzmann2011, damiano&hu2020, lin2020, Makherjee2021, palle2023}. These clouds, despite posing challenges in other techniques, enhance the potential of a positive detection in reflected light by increasing the planet’s overall reflectance.\\
\noindent Observing Earth as an exoplanet provides an essential benchmark for future rocky planet studies. Techniques such as spacecraft observations, disk-integrated spectra from satellites, and Earthshine observations offer crucial insights into how Earth-like planets might appear in reflected light \citep{robinson2018}. Contrary to satellite observations, Earthshine replicates the complex scattering and reflection geometry of how rocky exoplanets will be observed in reflected light. In recent years, Earthshine observations have been performed both in intensity \cite{goode2001, woolf2002, palle2003, tinetti2006a, palle2008, palle2016} and polarization \citep{sterzik2012, takahashi2013, bazzon2013, miles-paez2014, sterzik2019, sterzik2020, takahashi2021}. For ground-based observations, spectropolarimetry is particularly beneficial in distinguishing between telluric contamination, caused by Earth's own atmospheric transmission, and the actual reflected light from Earth. In standard spectroscopic observations, telluric lines overlap with the reflected light signals that we aim to detect, making it challenging to isolate the true reflected light features. Polarization filters out telluric lines, which remain largely unaffected by polarization, while retaining the polarized signal from the target.\\
\noindent The biggest challenge to observe rocky exoplanets in the habitable zone is the huge contrast between the stellar and planetary flux. Since F, G, and K-type stars emit nearly unpolarized light \citep{cotton2017}, polarization serves as a valuable tool for distinguishing planetary signals from stellar radiation. Incoming stellar light gets partially linearly polarized when reflected due to scattering in the atmosphere or surface reflections. Polarization has the potential to not only aid the detection of exoplanets, but also provides crucial insights into surface properties, enabling the differentiation between land and ocean surfaces.\\
\noindent Advanced radiative transfer models enable realistic simulations of Earth as an exoplanet, considering both intensity and polarization. \cite{stam2008} developed the first of such models using an adding-doubling approach, assessing the influence of the phase angle (i.e., the angle between the star, the planet, and the observer) and the wavelength dependence on the degree of linear polarization of various homogeneous planets. \cite{karalidi2012} extended the model to simulate inhomogeneous planets by dividing the planet into small pixels with homogeneous characteristics, finding that results differ significantly when compared with horizontally homogeneous exoplanets. This highlights how horizontal inhomogeneities leave distinct traces in flux and polarization signals. A more realistic setup for comparison with Earthshine observations was developed by \cite{emde2017}, using a 3D Monte Carlo radiative transfer code that accounts for the treatment of fully inhomogeneous spherical geometry, in contrast to plane-parallel, independent column models. The ability to simulate a realistic Earth-like exoplanet was of fundamental importance in assessing the influence of water and ice clouds, aerosols, and ocean glint on the polarization spectra of Earthshine and in constraining several biomarkers of our planet. In particular, \cite{emde2017} demonstrated that light reflected by the ocean surface in the ocean glint region causes the highest increase in polarization. \cite{trees2019} further improved the adding-doubling algorithm from \cite{stam2008} by incorporating a more realistic treatment of the ocean surface, including Fresnel reflecting waves and scattering within the water body.\\
\noindent Notably, the ocean glint feature enhances intensity and contributes strongly to linear polarization in both the planetary spectrum and the phase curve. The color change of the polarized flux appears to uniquely identify an ocean surface, independent of surface pressure or cloud fraction. This feature may serve as a key indicator of surface water. Model simulations by \cite{trees2022} predict that an ocean on the surface of an exoplanet might already be detectable through a spectropolarimetric measurement at a single phase angle: the glint leaves a dip instead of a peak in the polarization spectrum across the H$_2$O-gas absorption band around 950~nm. Additionally, using polarized rather than unpolarized reflectance makes it possible to exclude false-positive detections, which might be generated by reflecting dry surfaces or ice caps \citep{cowan2012}. However, multiple scattering by clouds and aerosols can depolarize the signal and increase the total flux, complicating interpretations \citep{hansen1971, emde2017}.\\
\noindent Polarization also plays a significant role in detecting liquid water clouds, as demonstrated by the presence of the polarized cloudbow feature \citep{karalidi2012_rainbow}. Even partial ice cloud cover does not completely obscure this signal. Observationally, \cite{sterzik2020} derived from spectropolarimetric Earthshine observations that clouds on Earth are made up of liquid water, inferring their particle size from the cloudbow feature as a demonstration of this technique for Earth-like exoplanets. This shows that spectropolarimetry can identify the composition of clouds.\\
\noindent In this first paper of a series, we present an updated modeling framework based on the 3D Monte Carlo radiative transfer code MYSTIC \citep{mayer2009, emde2017}, part of the libRadtran software package \citep{mayer2005, emde2016}, to simulate the Earth as an exoplanet. MYSTIC incorporates 3D Earth-like atmospheres, full Stokes vectors to account for polarization, and 2D inhomogeneous surface models. We further improve the surface modeling setup adding 2D inhomogeneous planetary surfaces, which can now couple the surface reflectance matrix for the ocean with Lambertian albedo maps for land surfaces, and inhomogeneous surface wind maps over the ocean, which improve the accuracy of the simulation of the ocean glint feature. Additionally, we developed a new 3D cloud generator algorithm to take into account sub-pixels cloud inhomogeneities with variable water content, using data from ERA5, the European Centre for Medium-Range Weather Forecasts (ECMWF) ReAnalysis 5th generation product, and satellite observations.\\
\noindent Finally, we conduct sensitivity studies to assess how these improvements impact the planet’s computed spectra and phase curves, highlighting the pronounced differences between true fine-scale surface and cloud structures versus simplified, coarse, and smeared-out models in observed intensity and polarization spectra. In the second paper of this series, we will apply our novel methodology to Earth-like exoplanets and explore its implications for a putative Earth 2.0 observed in reflected and polarized light. The third and final paper will compare our updated model with our entire catalog of Earthshine observations, validating our approach and establishing a baseline for using our models as "ground-truth" prescriptions in future studies on rocky exoplanet characterization and habitability assessment.

\section{Theoretical background}
\label{sec:theory}

Observing an exoplanet in reflected light, the contrast between the planet and the star can be expressed as
\begin{equation}
    \frac{F_p}{F_{\star}} = \left(\frac{R_p}{a}\right)^2 A_g \cdot g(\alpha) ,
    \label{eq:contrast}
\end{equation}
where $F_p$ and $F_{\star}$ are the fluxes of the planet and the star, respectively, $R_p$ is the radius of the planet, $a$ is the semi-major axis of the star-planet system, $A_g$ is the geometric albedo of the planet and $g (\alpha$) is the phase function, and $\alpha$ the phase angle (e.g., the angle between the star, the planet, and the observer). A reflected light spectrum of an exoplanet appears like the stellar spectrum with additional absorption and Rayleigh scattering features.\\
\noindent The geometric albedo is a wavelength-dependent quantity, while the phase function depends on both the wavelength and the phase angle. Both quantities are independent of the stellar spectrum.\\
\noindent By fixing the radius of the planet and the semi-major axis, we can express the contrast in fluxes only as a function of $A_g$ and $g(\alpha)$. This way, we can study the properties of different surfaces and clouds in the reflected light spectrum of an Earth-like planet.
We define this quantity as the reflectance, which is the product of the geometric albedo and the phase function. Thus the reflectance
$R = A_g \cdot g(\alpha)$ is also independent of the stellar spectrum.\\
\noindent The incident starlight to the planet is expected to be almost unpolarized. The disk-integrated sunlight is polarized at the level of 10$^{-6}$ \citep{kemp1987}. For active FGK-stars, the degree of linear polarization was calculated to be on the order of 23.0 $\pm$ 2.2 ppm, while for inactive stars the polarization signal is expected to be about 2.9 $\pm$ 1.9 ppm \citep{cotton2017}. Also, stellar flares and spots should contribute with small degrees of linear polarization on the order of $10^{-6}$ \citep{berdyugina2011}.\\
\noindent By contrast, stellar light reflected by the surface of the planet or scattered by processes happening in the exoplanet atmosphere can be linearly polarized at the level of several tens of percent. Rayleigh scattering produced by molecules in the atmosphere polarizes light, although multiple scattering processes involving cloud and aerosol particles can strongly depolarize the radiation. Similarly, certain surface features, like the ocean glint, produce strong linear polarization signatures, while scattering by other surface types can depolarize light previously polarized by Rayleigh scattering. Thus, observations of polarized light enhance the contrast between the planet and stellar fluxes.\\
\noindent We can describe the total, polarized and unpolarized, flux of the disk-integrated planetary signal with a Stokes vector \citep{Chandrasekhar1950}
\begin{equation}
    F = [I, Q, U, V],
\end{equation}
where $I$ is the intensity, $Q$ and $U$ are the linearly polarized fluxes and $V$ the circularly polarized flux. $Q$ and $U$ are described with respect to a reference frame. In our simulations, the reference plane corresponds to the planetary scattering plane, which is the plane through the centers of the planet, the host star and the observer. \\
\noindent We define the degree of polarization
\begin{equation}
    P = \frac{\sqrt{Q^2 + U^2 + V^2}}{I},
\end{equation}
which describes the fraction of photons which get polarized over the total flux coming from the planet, and it is independent on the reference plane. For an Earth-like atmosphere, $V$, the circular polarization component, is very small compared to the linear polarization terms, and thus we can express the degree of polarization 
\begin{equation}
    P = \frac{\sqrt{Q^2 + U^2}}{I}
    \label{degree_of_pol}
\end{equation}
only as a function of the linearly polarized components, thus becoming the degree of linear polarization.\\
\noindent While the total flux intensity spectrum of an exoplanet is an absolute measurement, the degree of linear polarization is a relative measurement and does not need to be calibrated with the distance or type of the star. Additionally, only the first few scattering orders contribute to the polarization signal. Multiple scattering depolarizes light because higher-order scattering events disrupt the predominant direction of radiation, reducing their contribution to the Stokes parameters $Q$ and $U$. In contrast, multiple scattering primarily contributes to $I$, thereby decreasing the degree of polarization $P$. Thus, the features in $Q$ and $U$ are better constrained compared to unpolarized light, because they are less smeared out by multiple scattering. For example, features like the cloudbow are due to single-scattering.\\

\section{3D Radiative Transfer simulations}
\label{3D_rad_tran}

We perform our simulations with MYSTIC \citep{mayer2009}, the Monte Carlo code for the phYsically correct Tracing of photons in Cloudy atmospheres, a versatile radiative transfer model for Earth's and planetary atmospheres in the libRadtran software package (\url{http://www.libradtran.org/doku.php}) \citep{mayer2005, emde2016}. The code is used here to calculate reflected and polarized light radiances of the Earth with realistic 3D atmospheres and 2D surface reflectance properties maps. The implementation of polarization in MYSTIC is described in \citet{emde2010}. Accurate and efficient cloud scattering simulations require the sophisticated variance reduction methods by \citet{buas2011}. The absorption lines importance sampling method \citep{emde2011} enables high spectral resolution simulations. In order to calculate the radiance reflected by the whole planet, a 3D spherical model setup is required \citep{emde2017}. The polarization results of MYSTIC show good agreement compared with other codes \citep{emde2015, emde2018, korkin2022}. \\
\noindent Here we summarize the major initial setup to generate spatially unresolved spectra of an Earth-like exoplanet:
\begin{itemize}
    \item The Sun-Earth-Moon geometry (e.g., phase angle) is simulated by selecting the Sun's and Moon's latitudes and longitudes relative to the Earth, along with the Earth-Moon distance. These values can be obtained from, e.g., \url{https://www.fourmilab.ch/earthview/};
    \item a sensor is placed at a distance $d = 384\,400~$km, the typical Earth-Moon distance, with an aperture $\theta = 1.25$\degr.  As a result, the scene consists of the Earth at the center of the field of view, surrounded by empty space. The reflectance values produced by MYSTIC are artificially reduced due to photons traveling into empty space without being reflected back. To correct for this effect, all reflectance values are divided by a geometrical factor:
    \begin{equation}
        f = \frac{(2d\tan{\theta})^2}{\pi R_{\oplus}^2}
    \end{equation}
    where $R_{\oplus}$ is the Earth radius;
    \item we run $10^7$ photons to generate each spectrum, resulting in a standard deviation below 0.1\% \citep{emde2017}, which is significantly smaller than the expected observational errors. For the single-wavelength simulations used in phase curve calculations, we run $10^6$ photons, reaching a relative error of roughly 0.3\%. To generate the images, we divide each image into a 1000 × 1000 pixel grid and run 1000 photons per pixel, resulting in a total of $10^9$ photons;
    \item all spectra are obtained with a spectral resolution of 1~nm over the 400–1000~nm wavelength range. The phase curves are computed with an angular resolution of 2\degr. While the spectra are calculated in the optical wavelength regime, they can also be computed in the ultraviolet (UV), near-infrared (NIR), and mid-infrared (MIR) regimes. Additionally, the spectral resolution can be adjusted to achieve high-spectral resolutions of up to $R$ = 100000;
    \item an atmosphere is specified by a vertical pressure and temperature profile, as well as the vertical mixing ratios of the most common gas species. We use the US standard amtosphere properties \citep{anderson1986};
    \item 3D ice and liquid water clouds are included from a cloud file which specifies the 3D spatial position, liquid water content, ice water content, and effective radius of cloud droplets and ice crystals. The optical properties of liquid water clouds are computed using the Mie scattering tool of the libRadtran package \citep{emde2016, wiscombe1980}. The ice crystals are composed of smooth randomly oriented crystals treated as a mixture of six habits, and their optical properties are parametrized following the HEY (Hong, Emde, Yang) parameterization, which includes the full phase matrices for the 0.2 to 5~$\mu$m wavelength range \citep[Appendix A]{emde2016};
    \item we generate realistic cloud distributions from the ECMWF ERA5 data for a particular date and time, and in this paper we introduce a new 3D Cloud Generator approach (Sec. \ref{sec:3DCG}) to account for cloud inhomogeneities and sub-grid variability;
    \item land surfaces are treated as Lambertian surfaces with different albedos, and in this paper we introduce new wavelength-dependent surface albedo maps \citep{Roccetti2024} in 3D radiative transfer simulations (Sec. \ref{sec:albedo_maps}) to treat spectral features of different land components;
    \item the ocean surface is simulated using the bidirectional surface reflection functions (BPDFs) coupled with a surface wind speed maps acquired from ERA5 data and accounts for the influence of waves, including shadowing effects \citep{Mishchenko1997}.
\end{itemize}
\noindent \cite{emde2017} used MYSTIC to simulate the Earth as an exoplanet and compare the results with Earthshine observations. We build upon the approach of \cite{emde2017} by greatly enhancing the representation of surface properties. While \cite{emde2017} could only simulate either an ocean surface with its BPDF or a purely Lambertian surface, our novel method allows for a combination of both. It is now possible to represent inhomogeneous planetary surfaces coupling an ocean surface with a Lambertian representation of land, which greatly improves the realism of surface modeling. Additionally, we now include inhomogeneous surface wind maps provided by the ERA5 reanalysis products for the treatment of surface wind speed, expanding from the homogeneous wind maps available in \cite{emde2017}.\\
\noindent A second, most important, improvement is the refinement of the 3D cloud profile implementation. We perform sensitivity studies to address the need of introducing cloud inhomogeneities and sub-grid variability, which reduces the bias from the former cloud representation (\cite{emde2017}, see Sec. \ref{sec:3DCG} for more details). Lastly, we introduce variable effective cloud droplet radii based on realistic parametrisations from ECMWF \citep{ECMWF2024}.\\
\noindent To generate true color images of the planet, we first perform MYSTIC simulations at three wavelengths: 469~nm, 555~nm and 645~nm. From these three wavelengths, a simplified spectrum is constructed as a step function in the visible range, where intensity values at all wavelenghts in the continuum are associated with the output of the closest simulated wavelength. The simplified spectrum is then convoluted with the color matching functions from the International Commission on Illumination (CIE) to obtain colors in the CIE color space. CIE colors are then converted to RGB colors using the transformation of the RGB standard. The resulting colors have a linear brightness scale and need to be transformed into a power-law brightness scale by applying a gamma transfer function, as specified by the RGB standard.

\section{Surface modeling}
\label{sec:surface}

Ocean and land surfaces exhibit distinct albedo characteristics due to their differing properties. Oceans have low albedo because water absorbs most visible and near-infrared sunlight. In contrast, land surfaces vary in reflectance based on composition: snow-covered regions are highly reflective, while forests and deserts show lower albedo depending on their vegetation and soil type. Surface roughness further influences reflectance. Smooth surfaces, like calm oceans or bare rock, reflect more light than uneven surfaces, such as forests and grasslands. Additionally, the angle of incoming sunlight affects reflectance. Sunlight striking the ocean surface at a steep angle tends to penetrate the water, while shallow-angle sunlight is more likely to be reflected.\\
\noindent In \cite{emde2017}, simulating both ocean and Lambertian land surfaces simultaneously was not possible in polarization mode. In our updated approach, we now implement a Lambertian surface for land pixels, specifying albedo values, and a BPDF for ocean pixels. This is the only model available in the literature that can effectively handle a mixture of BPDF and Lambertian surfaces, compared to previous work by \cite{emde2017, trees2019, trees2022, vaughan2023}. This enables realistic simulations of inhomogeneous planetary surfaces, combining ocean and land, and improving the treatment of varying albedo effects for Earth and exoplanets.\\
\noindent All land surface types, apart from oceans, are treated as pure Lambertian surfaces in our model. Although materials like forest leaves and sand can polarize light via surface reflection, land BPDFs are unavailable for our model. However, the largest polarized reflectances observed by the PARASOL instrument over land are of the order of 0.02 and 0.04 \citep{Maignan2009}, and their contribution to Earth’s disk integrated polarized radiance is expected to be minor, as shown in \cite{groot2020}.\\
\noindent In the case of water reflections, surface roughness significantly expands the area where sunlight is directly reflected, extending beyond the solar disk's size on Earth. This phenomenon, known as ocean glint, becomes more pronounced with increasing wind speed over the water surface. For our simulations, the ocean is modeled using the realistic BPDF from \cite{Mishchenko1997}. This model also incorporates the shadowing effects of ocean waves, as described by \cite{tsang1985}.

\subsection{Hyperspectral albedo maps}
\label{sec:albedo_maps}

\begin{figure*}[h]
    \includegraphics[width=1\linewidth]{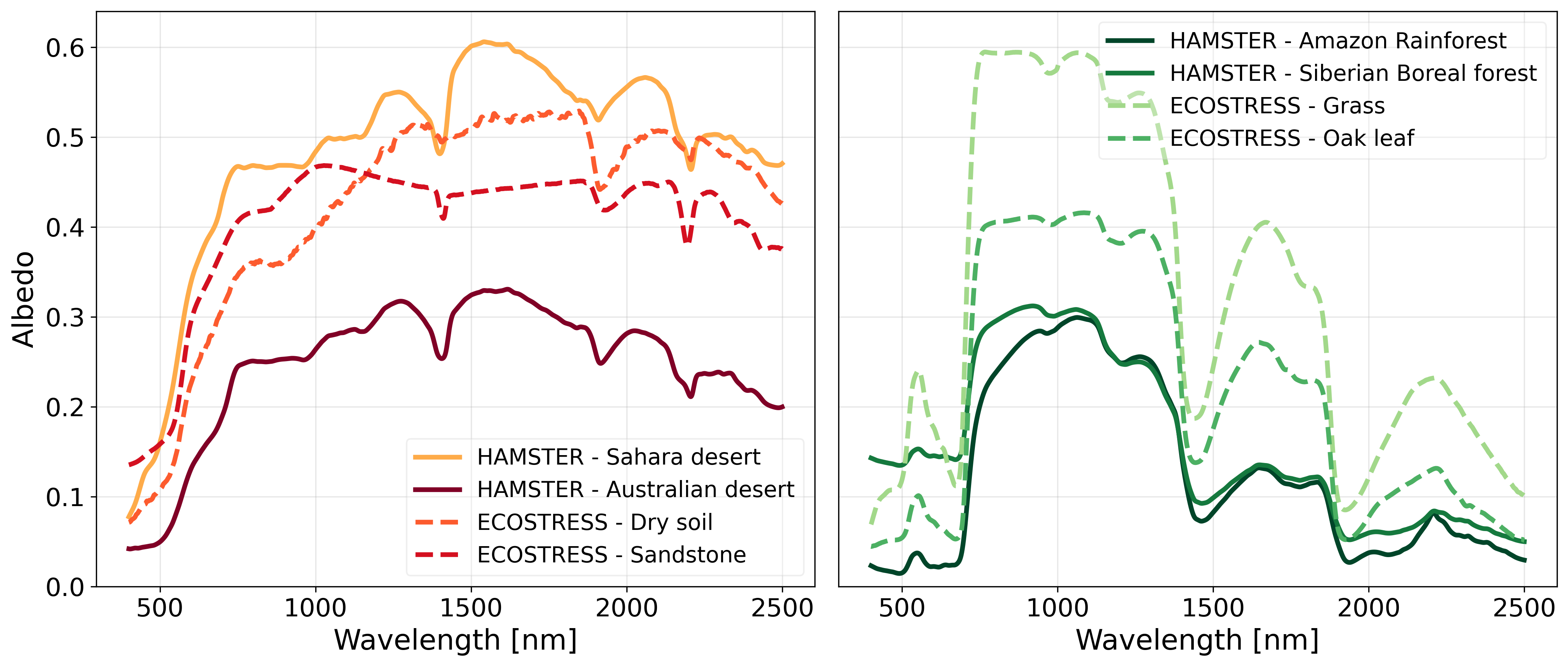}
    \caption{Comparison of desert and forest spectra from HAMSTER (solid lines) with typical soil and leaf spectra from the ECOSTRESS library (dashed lines).}
    \label{fig:VRE}
\end{figure*}

Albedo varies with wavelength, and accurately representing this spectral variability is crucial for correct energy balance calculations in radiative transfer simulations. Proper treatment of the spectral variability of surface types is essential for capturing the continuum in simulated spectra and identifying features like the VRE around 780~nm. However, no global hyperspectral albedo maps are provided by current satellite or reanalysis products.\\
\noindent \cite{gordon2023} addressed this issue by classifying surface pixels into five categories (ocean, forest, grass, sand, and snow/ice) using the Level-3 MODIS Yearly Global Land Cover Types (YGLCT) product. They coupled these categories with the ECOSTRESS spectral library \citep{Baldrige2009, Meerdink2019} to account for spectral variability. A similar approach was used by \cite{Kofman2024}, who applied linear mixing of land surface spectra to represent mixed surface types.\\
\noindent \cite{Roccetti2024} developed a hyperspectral albedo map dataset, HAMSTER, which captures the spatial, temporal, and spectral variability of Earth's surface albedo, based on MODIS surface albedo satellite measurements \citep{Schaaf2002}. This dataset significantly improves land surface representation by treating them as a mixture of soils and vegetation. In terms of seasonal variability, HAMSTER shows that rainforests exhibit almost no seasonal changes, whereas boreal forests experience variations due to snow, with their summer profile closely resembling that of a rainforest. Similarly, for deserts, HAMSTER reveals that the Sahara and Australian deserts show little to no seasonal variation, while high-altitude deserts are affected by seasonal snow cover. In contrast, the typical profile of a savanna is strongly season-dependent, resembling a desert during the dry season and a forest during the wet season \citep{Roccetti2024}.\\
\noindent In Fig. \ref{fig:VRE}, we compare the typical reflectance spectra of different soils and vegetation components from the ECOSTRESS library with the yearly average, spatially integrated spectra of a forest and a desert from HAMSTER \citep{Roccetti2024}. Specifically, we compare the Sahara and Australian deserts in ECOSTRESS with dry soil and sandstone materials. Similarly, we compare the Amazon rainforest and the Siberian boreal forest with the grass and oak leaf spectra from ECOSTRESS.\\
\noindent We find that without such detailed treatment, land reflectance can be over- or underestimated. For instance, the VRE peak at 780~nm reaches around 0.3 in \cite{Roccetti2024}, whereas ECOSTRESS data for a broadleaf tree shows an albedo of approximately 0.6, underscoring the importance of combining soil and vegetation properties in land surface simulations. A similar trend is observed for the "green bump" around 500-600~nm, which is also overestimated when the combination of various surface types is not considered.\\
\noindent We use HAMSTER to represent surface reflectance in our model and explore the sensitivity of surface albedo to seasonal variability in Sec. \ref{sec:albedo_results}. HAMSTER provides a spatial resolution of 0.05\degr in latitude and longitude, a temporal resolution of one day, and a spectral resolution of 10~nm, allowing it to capture key spectral features such as chlorophyll absorption and other soil properties.\\
\noindent To evaluate the impact of HAMSTER, we replicated the simplified hyperspectral albedo maps described in \cite{Kofman2024}. This was achieved by generating a linear combination of spectra from the ECOSTRESS spectral library, guided by land cover type information from the MODIS MCD12C1 product. The 17 land cover types were grouped in five categories (grass, forest, soil, snow and water bodies) as in \cite{gordon2023} and \cite{Kofman2024}. The resulting spectra and phase curves were then compared to assess their differences.

\subsection{Inhomogeneous surface wind maps}
\label{sec:wind}

The reflective properties of the ocean glint depend on the water surface structure, which is influenced by wind speed. In our model, we use horizontal surface wind data from meteorological sources as input to the BPDF function for water surfaces \citep{tsang1985, Mishchenko1997}. For each surface grid cell, the wind speed is calculated from the northward and eastward wind components provided by the "ERA5 hourly data on single levels from 1940 to present" \citep{Hersbach2020}. This approach generates inhomogeneous surface wind maps that affect the shape and size of the glint, although the wind direction is not considered. This is the first time that inhomogeneous surface wind speed maps are introduced, while previous approaches used a constant wind speed of 10\,m\,s$^{-1}$ \citep{emde2017}, and between 1 to 13\,m\,s$^{-1}$ \citep{trees2022}. This setup provides a realistic representation of the glint, which, together with cloud cover, impacts the spectra and phase curves of the planet. The relationship between ocean glint and wind speed is explored in Section \ref{sec:appendix_D}.

\section{Cloud modeling}
\label{sec:clouds}

We introduce a novel approach to modeling clouds in our 3D simulations to achieve a highly realistic representation of global cloud systems on Earth. As the detailed combination of clouds and surfaces determines the geometrical albedo, accurate prescriptions are essential to model reflectance and polarization correctly (Eq. \ref{eq:contrast}). In our model, we represent clouds using 3D maps of liquid water content ($LWC$) and ice water content ($IWC$) from the ECMWF's ERA5 reanalysis product \citep{Hersbach2020}. Reanalysis products combine past observations with advanced weather forecast models to generate a consistent and complete picture of past atmospheric conditions. This involves the use of data assimilation techniques to integrate observations from various sources (such as satellites, weather stations, and buoys) into a weather model. The ERA5 reanalysis product comes with a high horizontal spatial resolution of approximately 28~km on a global grid (or 0.25\degr) and 37 vertical levels from the surface up to 1 hPa (about 48~km altitude), providing detailed information on the state of the atmosphere. The $LWC$ and $IWC$ are the starting point to calculate the optical thickness of clouds. In the approximation of geometrical optics, which is accurate in the visible, the optical thickness is given by
\begin{equation}
    \tau = \frac{3}{2} \frac{\Delta H \cdot LWC}{\rho_{\rm w} \cdot r_{\rm eff}},
\label{eq:tau}
\end{equation}
where $\Delta H$ is the height of the cloud, $\rho_{\rm w}$ is the density of liquid water and $r_{\rm eff}$ is the cloud droplet effective radius. For an ice water cloud, $\tau$ is calculated as in Eq. \ref{eq:tau} by substituting $LWC$ with $IWC$. The ERA5 data comes with mass mixing ratios which are converted to $LWC$/$IWC$ by computing the local air density using the ideal gas law.\\
\noindent ERA5 provides hourly data, allowing for detailed temporal analysis of weather and climate patterns. This high temporal resolution is valuable for studying diurnal cycles, extreme weather events, and other time-sensitive phenomena. The consistency of the ERA5 reanalysis product is crucial for long-term climate studies, trend analysis, and the validation of climate models. \\
\noindent To model a representative planet, we run all our cloudy spectra and phase curves with 12 different cloud fields selecting one random day from each month during 2023. This way, we can represent the variability of the planet, like its seasonal and internal variability. We select different random days for the two different geometries we are representing: 
\begin{itemize}
    \item for an Ocean planet, defined as a scenery over the Pacific Ocean, we simulate cloudy planets of the days reported in Table \ref{tab:Ocean}, at UTC 22:00 to catch the cloud properties as they are over the Pacific ocean, better representing an Ocean planet configuration;
    \item for an Earth-like planet, we select a scenery over the Indian ocean, with Asia, Europe and Africa visible in the scene. The cloud fields are randomly selected by day, but are all selected for UTC 06:00, to better represent cloud properties which are a mixture of land and ocean components. This prevents introducing a bias associated with cloud properties over different surfaces. The cloud properties for the Earth-like planet are displayed in Table \ref{tab:Earth-like}.
\end{itemize}
When comparing cloud properties over the Ocean and the Earth-like planet, we observe small differences in cloud cover, optical thickness, and the effective radii of water and ice cloud droplets. Throughout the study, we represent the variability due to cloud properties in all spectra and phase curves by averaging the final spectra across 12 independent simulations, each using different cloud fields, and indicating the 1$\sigma$ spread resulting from cloud variability. \\
\noindent We also notice that the cloud cover value we find from the ERA5 reanalysis cloud product is rather large. A possible explanation is the "resolution effect" problem. \cite{dutta2020} found the total cloud cover measured by finite resolution satellites to be overestimated. This bias can be overcome with finer angular resolution satellite instruments.\\
\noindent \cite{emde2017} used the predecessor of ERA5, ERA Interim, to represent cloud properties within MYSTIC. ERA Interim was provided on a coarser grid than the ERA5 product. Using hourly data such as the ones provided by ERA5 is useful to compare with Earthshine observations, as it provides the opportunity to represent the cloud properties at the exact date and time of the observations. However, \cite{emde2017} assumed that the $LWC$ and $IWC$ provided by ERA Interim were the in-cloud values, thus they mutiplied them by the cloud fraction of each gridbox to obtain the gridbox average $LWC$ and $IWC$, which are the quantities needed to estimate the cloud optical thickness in MYSTIC. However, as in ERA5, also ERA Interim was providing already the gridbox averages $LWC$ and $IWC$, thus the multiplication for the cloud fraction substantialy underestimated the cloud optical thickess and impacted the representation of the radiative and microphysical properties of clouds. This effect has significant impact and is the major reason why the Earthshine polarization spectra modeled in \cite{emde2017} did not match well observations.\\
\noindent While ERA5 presents a rather fine spatial resolution for a global cloud model, the grid size of each pixel (around 31~km) is still rather large and fails to represent the more realistic, patchy nature of clouds. In fact, smoothing microphysical cloud properties over large grid scales results in an average state of clouds and spreads the $LWC$ and $IWC$ information on the entire grid cell. However, in nature, clouds appear to be patchy and a more correct representation is required to account for horizontal inhomogeneities among clouds and the presence of clear-sky and cloudy sub-pixels inside the same grid cell. This effect of cloud representation is very relevant in 3D radiative transfer models, and in particular for polarized spectra and phase curves, which are more sensitive on the effect of clouds and their microphysical properties. Thus we developed a new approach to treat 3D cloud sub-pixel variability and inhomogeneities, called the 3D Cloud Generator approach, in order to calculate reference (ground-truth) spectra and phase curves of Earth as an exoplanet in reflected and polarized light.\\

\subsection{3D Cloud Generator}
\label{sec:3DCG}
\begin{figure*}
    \centering
    \includegraphics[width=0.9\linewidth]{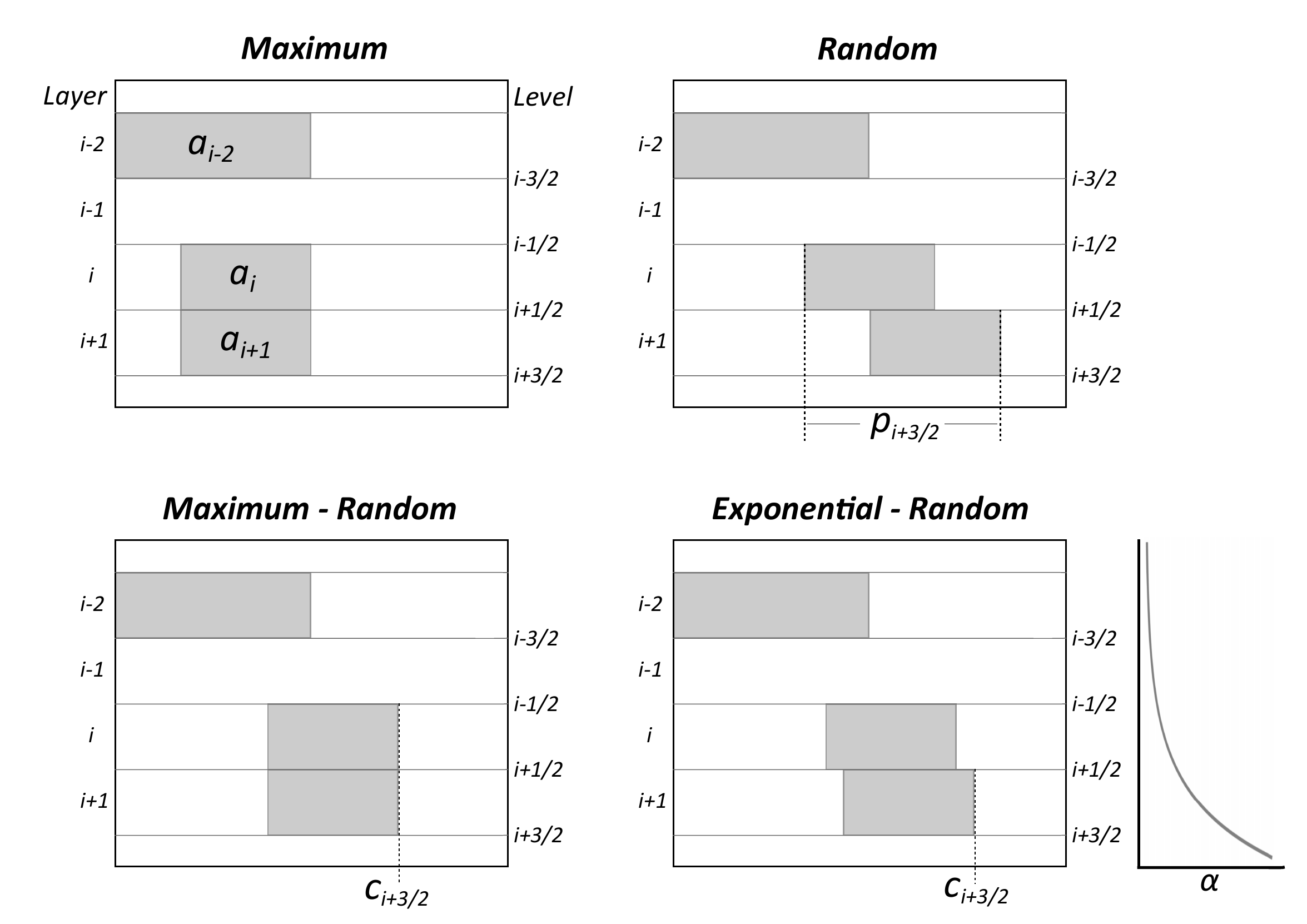}
    \caption{Schematic illustration of four cloud overlap assumptions: maximum, random, maximum-random (MAX-RAN) and exponential-random (EXP-RAN) overlap. The figure also shows the meaning of three important quantities: the cloud cover of a given layer, $a_i$, the pairwise cloud cover $p_{i+1/2}$ and the cumulative cloud cover $c_{i+1/2}$. For the EXP-RAN overlap, a schematic illustration of the exponentially decaying $\alpha$ parameter is also included, showing how it decreases with altitude, as the separation between layers increases.}
    \label{fig:sketch}
\end{figure*}

The 3D Cloud Generator (3D CG) is based on the \cite{hogan&bozzo2018} one-dimensional cloud generator, but adapted for a 3D radiation scheme. This is crucial to generate realistic spectra and phase curves of Earth as an exoplanet, due to the horizontal and vertical distribution of clouds, and in particular their overlap over the ocean glint area. Our 3D CG, being based on the one used from the 1D radiation scheme ecRAD, does not include any correlation between columns, which is left for a future implementation.\\
\noindent The first step in the 3D CG is to compute the cumulative cloud cover profile, $c_{i+1/2}$, from the top of the atmosphere (TOA) to the $i$-th layer, using ERA5 cloud cover data ($a_i$) for individual layers. Additionally, the pairwise cloud cover, $p_{i+1/2}$, between adjacent layers $i$ and $i+1$ is calculated. However, without sub-grid information on the horizontal cloud distribution, determining $p_{i+1/2}$ from $a_i$ and $a_{i+1}$ is not straightforward. Clouds can be vertically overlapping, randomly arranged or a mixture of both. To address this ambiguity, various cloud overlap assumptions have been developed and validated against observations \citep{Hogan2000}. In Figure \ref{fig:sketch} we show a schematic illustration of four cloud overlap assumptions.\\
\noindent The two most simple cloud overlap assumptions are the maximum and the random overlaps. The maximum overlap assumes clouds are perfectly vertically aligned, while the random overlap assumes a completely stochastic distribution between layers. These two assumptions represent opposite extremes, while the maximum-random (MAX-RAN) overlap \citep{Morcrette2000} offers a more balanced and realistic approach. In MAX-RAN overlap, clouds in vertically contiguous layers are maximally overlapped, while those separated by cloud-free layers are randomly overlapped.\\
\noindent Assuming MAX-RAN overlap, the pairwise cloud cover of two adjacent layers is:
\begin{equation}
    p_{i-1/2} = \max(a_{i-1}, a_i).
\end{equation}
An alternative approach is the exponential-random (EXP-RAN) overlap \citep{Hogan2000}, where an "overlap parameter" $\alpha$ is used to interpolate between the maximum and the random overlap assumptions for adjacent cloudy layers. The $\alpha$ parameter decays as the distance between layers increases:
\begin{equation}
    \alpha_i = \text{exp}\left({-\frac{\Delta z_i}{z_0}}\right),
    \label{eq:overlap_parameter}
\end{equation}
where $\Delta z_i$ is the thickness of the $i$-th layer and $z_0$ is the cloud cover decorrelation length. The ECMWF Integrated Forecast System (IFS) documentation \citep{ECMWF2024} presents a parametrization of $z_0$ based on latitude, following the ideas of \cite{Shonk2010}:
\begin{equation}
    z_0[\text{km}] = 0.75 + 2.149\cos^2{\phi}.
\end{equation}
The thickness of the layers, $\Delta z_i$, increases with altitude for ERA5 reanalysis data, hence the $\alpha$ parameter decreases exponentially with altitude.\\
The pairwise cloud cover computed assumping EXP-RAN overlap is:
\begin{equation}
    p_{i-1/2} = \alpha_{i-1/2}\max(a_{i-1}, a_i) + (1- \alpha_{i-1/2})(a_{i-1} + a_i - a_{i-1}a_i).
\end{equation}
where the $\alpha$ parameter weights more the maximum overlap for close layers and the random overlap for distant ones.\\
For both assumptions, we compute the cumulative cloud cover iteratively as:
\begin{equation}
    c_{i+1/2} = 1 - (1-c_{i-1/2})\frac{1-p_{i-1/2}}{1-a_{i-1}}.
\label{eq:exp_ran_cloud_cover}
\end{equation}
This first step is completely deterministic, and only depends on the cloud overlap assumption employed.\\
\begin{figure*}
    \centering
    \includegraphics[width=0.8\linewidth]{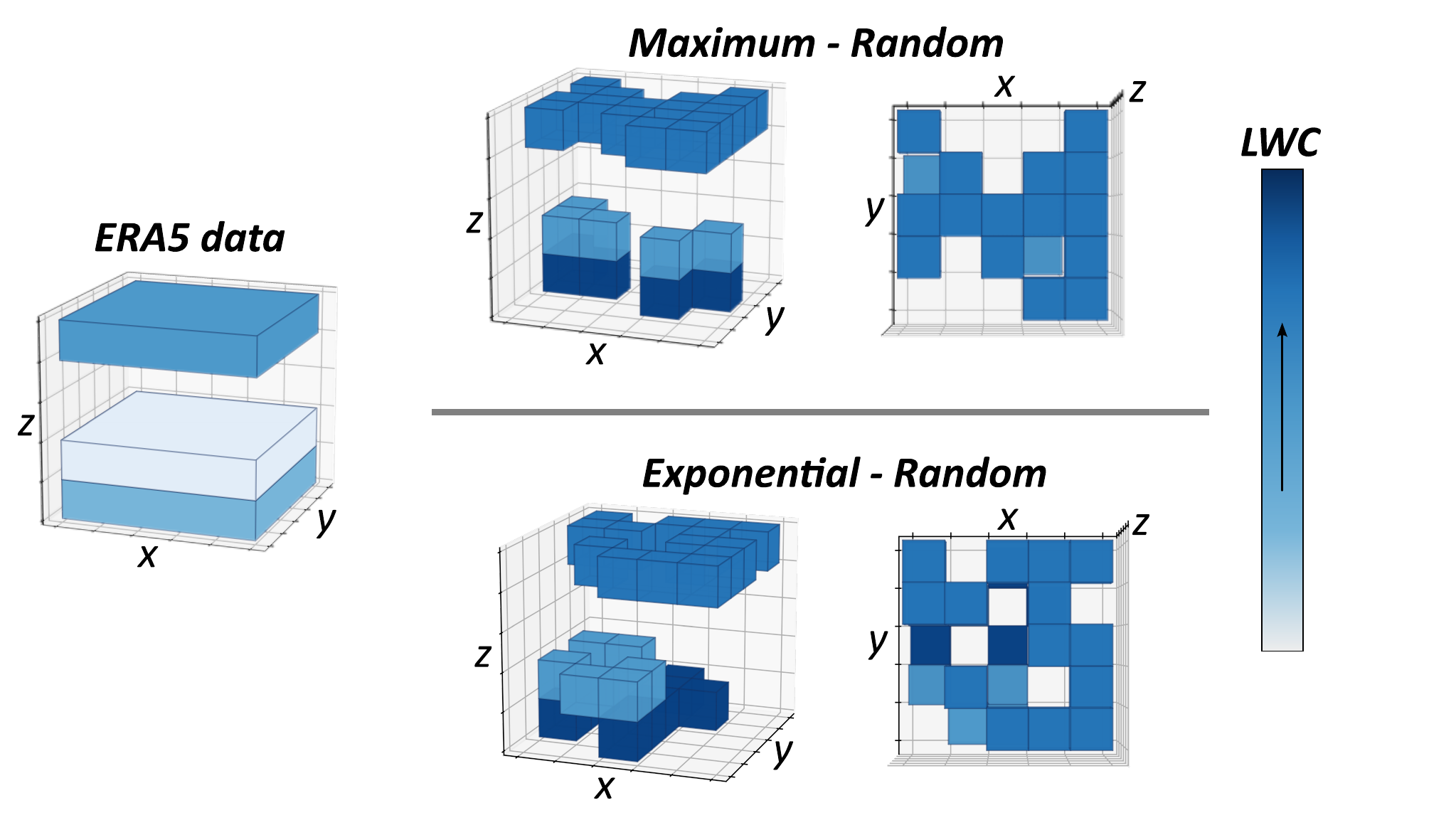}
    \caption{A 3D schematic illustration depicting the input and output of the 3D CG for liquid water clouds, without including horizontal inhomogeneity. The original ERA5 pixels are subdivided into smaller columns, with clouds assigned based on the procedure outlined in the main text. The color of the boxes represents the associated $LWC$, which is adjusted to conserve the ERA5 $LWC$ layer by layer.}
    \label{fig:3D_plot}
\end{figure*}
\noindent The next step of the 3D CG is to divide the original ERA5 column in $N\times N$ subcolumns and assign, along each subcolumn (independently) cloudy and non-cloudy sub-pixels. This procedure is stochastic and is again based on \cite{hogan&bozzo2018}. Initially, we determine the highest cloudy layer by generating a first random number $R_0$ from a uniform distribution between 0 and 1. The $i$-th layer is chosen as the highest cloudy one if $c_{i-1/2}/C < R_0 \leq c_{i+1/2}/C$, where $C=c_{n+1/2}$ is the total cloud cover. Next we proceed further down to the layer $i+1$ and determine if that layer is also cloudy by generating a further random number $R_i$. If the present layer $i$ is cloudy, the next one is cloudy too if:
\begin{equation}
    R_i < \frac{a_i+a_{i+1}-p_{i+1/2}}{a_i}.
\label{eq:prob_cloudy}
\end{equation}
On the other hand, if the current layer $i$ is not cloudy, the next one is set to be cloudy if:
\begin{equation}
    R_i < \frac{p_{i+1/2} - a_i - c_{i+3/2} + c_{i+1/2}}{c_{i+1/2} - a_i}.
\label{eq:prob_noncloudy}
\end{equation}
The right hand side of Eq. \ref{eq:prob_cloudy} (and Eq. \ref{eq:prob_noncloudy}) represents the conditional probability of finding a cloudy pixel in layer $i+1$ given that the previous layer $i$ was cloudy (or not). For a more thorough discussion on how this probability is derived, see Fig. 3 in \cite{hogan&bozzo2018} and the accompanying explanation.\\
\noindent By the end of this procedure, we obtain a binary matrix $\mathcal{C}$ with dimensions $(N,N,N_z)$, with $N_z$ the number of vertical layers, recording which pixels are populated by clouds and which are non-cloudy.\\
\noindent The last step of the 3D CG consists of assigning to all cloudy pixels a value for $LWC$ and $IWC$. Contrary to \cite{hogan&bozzo2018}, we start directly from the ERA5 gridbox averages $\overline{LWC}_i$ and $\overline{IWC}_i$ instead of scaling the optical depth.\\
\noindent First we convert the gridbox average $\overline{LWC}_i$ to the in-cloud $LWC_{i}^{ic}$ of layer $i$ by dividing by the cloud cover $a_i$: 
\begin{equation}
    LWC_{i}^{ic} = \frac{\overline{LWC}_i}{a_i}.
\end{equation}
The same holds for the in-cloud $IWC_{i}^{ic}$. In principle we could set all cloudy pixels in a given layer $i$ of the matrix $\mathcal{C}$ to this in-cloud $LWC_i^{ic}$ value. In practice, we follow the algorithm of \cite{Räisänen2004} to generate a distribution of $LWC$ that resembles more closely the cloud cover vertical distribution and to introduce some horizontal inhomogeneity. The uppermost cloudy layer $LWC_i$ is determined by scaling the $LWC_i^{ic}$ by a random value $\gamma$ extracted from a gamma distribution of mean value $\mu = 1$ and fractional standard deviation $\sigma = 0.75$. If the next layer $i+1$ is cloudy, we extract a random number $R_{i+1}$ and compare with the cloud condensate cumulative frequency $\alpha^{lwc}_{i+1}$. If $R_{i+1} < \alpha^{lwc}_{i+1}$, the next layer is filled with the same $LWC$ as the previous layer; if the opposite is true, a new random value from the gamma distribution is extracted. In summary:
\begin{equation}
    LWC_{i+1} = \begin{cases}
        LWC_{i} &\text{if    } R_{i+1} < \alpha^{lwc}_{i+1} \text{    and    } \mathcal{C}_i = 1\\
        \gamma \cdot LWC_{i+1}^{ic} &\text{otherwise}
    \end{cases}
\label{eq:LWC_gamma}
\end{equation}
The cloud condensate cumulative frequency $\alpha^{lwc}$ is computed as the exponential overlap parameter $\alpha$ (Eq. \ref{eq:overlap_parameter}), the only difference being that the decorrelation length $z_0$ of the cloud parameters is half the one for the cloud cover \citep{ECMWF2024}.\\
\noindent By the end of this procedure, the randomness intrinsic to the attribution of $LWC$ values may have changed the total amount of $LWC$ compared to the initial gridbox average data from ERA5. Therefore, we add an additional step compared to the original procedure by \cite{Räisänen2004} and we correct for this randomness by rescaling, layer by layer, each value of $LWC_i$ by:
\begin{equation}
    LWC_i^r = LWC_i \cdot \frac{LWC_i^{ic}}{<LWC_i>}
\label{eq:LWC_scale}
\end{equation}
where the mean $<LWC_i>$ is performed over all sub-columns and $LWC_i^r$ is the rescaled $LWC$ of the layer $i$. The same procedure is performed also to obtain the values for the $IWC$.\\
\noindent The 3D CG input and output are shown in Fig. \ref{fig:3D_plot}. The original ERA5 pixel is divided into subcolumns, which are populated by cloudy and non-cloudy pixels according to the procedure described above. Additionally, $LWC$ and $IWC$ values for all pixels are obtained using Eq. \ref{eq:LWC_gamma} and Eq. \ref{eq:LWC_scale}. As shown in the figure, the EXP-RAN overlap scheme generates slightly more cloudy columns, as it introduces randomness even in contiguous cloud layers, unlike the MAX-RAN overlap. 

\subsection{Variable cloud effective radius}
\label{sec:eff_radius}

The optical thickness depends also on the effective radius of the cloud droplets as in Eq. \ref{eq:tau}. $r_{\rm eff}$ is calculated following the ECMWF parametrizations \citep{ECMWF2024}. The liquid water effective radius is:
 \begin{equation}
     r_{\rm eff}^{\rm liquid} = \left[\frac{3E_d(LWC+RWC)}{4\pi\rho_{\rm w} k N_d}\right]^{1/3},
 \end{equation}
where $RWC$ is the rain water content, $E_d$ is an enhancement factor accounting for drizzle dispersion introduced in \cite{Wood2000}, $k$ is a shape-dependent factor and $N_d$ is the number concentration of cloud droplets, parametrized differently between pixels above sea and pixels above land following \cite{Martin1994} and computed using the wind speed as an input.\\
\noindent The ice effective radius is parametrized following \cite{Sun1999} and \cite{Sun2001}:
\begin{equation}
    r_{\rm eff}^{\rm ice} = \frac{3\sqrt{3}}{8} D_{\rm eff}^{\rm ice}, 
\end{equation}
where $D_{\rm eff}^{\rm ice}$ is the ice particle effective diameter. For all the details regarding these parametrizations, we refer to \cite{ECMWF2024}.\\
\noindent We also investigate the effect of using a constant effective radius, instead of a variable effective radius calculated for each gridbox from the ECMWF parametrization, while keeping the same optical depth. This requires us to rescale the $LWC$ of the cell accordingly, given the equation for the optical depth Eq. \ref{eq:tau}:
\begin{equation}
    LWC' = LWC_0 \cdot \left(\frac{r_{\rm eff}'}{r_{\rm eff,0}}\right),
\end{equation}
where $LWC_0$ and $r_{\rm eff,0}$ refer to the standard quantities, using the parametrized effective radius, and $LWC'$ and $r_{\rm eff}'$ represent the variables in the case of constant effective radius.

\subsection{Cloud zoom-out algorithm}
\label{sec:zoom-out}

To further investigate the effect of the horizontal grid resolution in simulating spectra and phase curves of spatially unresolved exoplanets, we now move to coarser grid scales. We call this procedure cloud zoom-out, as we average out the cloud properties from the fine grid scales of the 3D CG and the ERA5 product to coarser resolutions.
To generate 3D cloud maps at coarser resolutions, we average out the ERA5 data across $N\times N$ horizontal grid-cells, while keeping the vertical direction intact and the optical depth constant. In order to do so, we first compute the average ratio between $LWC$ and $r_{\rm eff}$ and then we multiply it by the average $r_{\rm eff}$. In this way, we keep the optical depth of the new, merged, column, equal to the mean of the original sub-columns on which we performed the averaging.
\begin{align}
&\overline{r_{\rm eff}} = \langle r_{\rm eff} \rangle\\
&\overline{LWC} = \biggl\langle\frac{LWC}{r_{\rm eff}} \biggr\rangle \times \langle r_{\rm eff}\rangle , 
\end{align}
where the brakets indicate the average over the $N\times N$ horizontal window. $\overline{r_{\rm eff}}$ and $\overline{LWC}$ are the values assigned to the averaged column.

\section{Results}
\label{sec:results}

\begin{figure*}[h]
    \centering
    \includegraphics[width=1\linewidth]{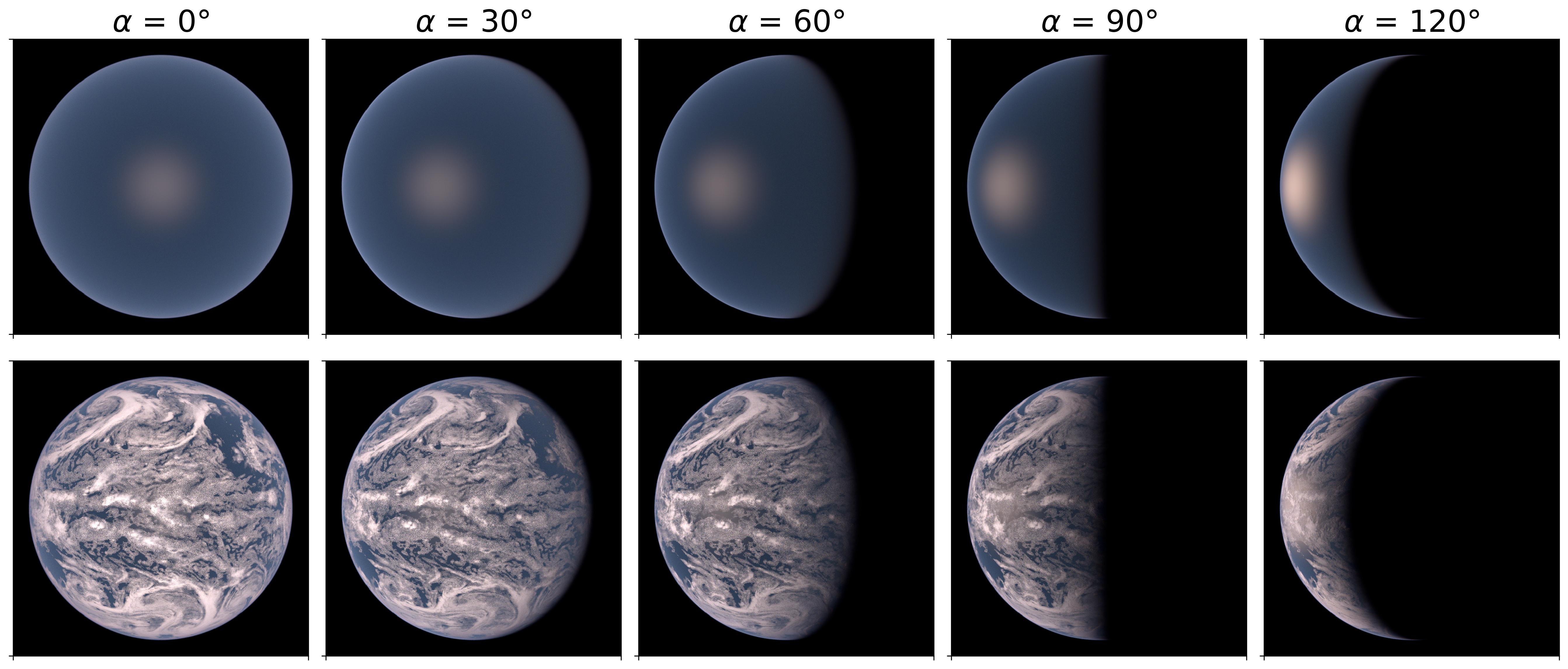}
    \caption{True colour simulations of the Ocean planet configuration with an Earth-like atmosphere for different phase angles (different colums). The same geometry is simulated without (first row), and with clouds (second row), using the ERA5 reanalysis dataset from 2023-09-18 UT22:00 for the cloud setup. The glint feature becomes covered by inhomogenous realistic clouds in the second row.}
    \label{fig:images_ocean}
\end{figure*}

\begin{figure*}[h]
    \centering
    \includegraphics[width=1\textwidth]{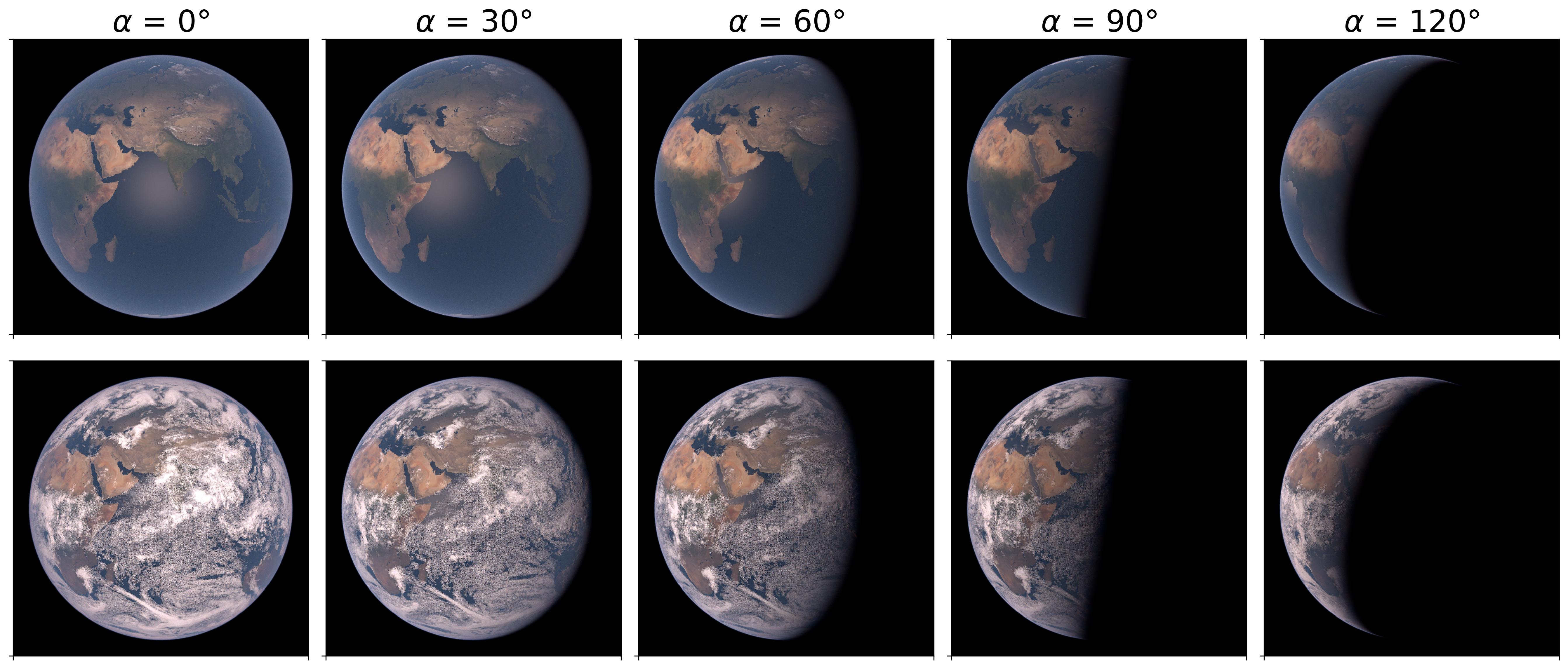}
    \caption{True colour simulations of the Atlantic configuration of an Earth-like exoplanet for different phase angles (different colums). The same geometry is simulated without (first row), and with clouds (second row), using the ERA5 reanalysis dataset from 2023-09-18 UT06:00 for the cloud setup. The glint feature becomes partially hidden by continents in the first row, and covered by inhomogenous realistic clouds in the second row.}
    \label{fig:images_earth}
\end{figure*}

Using MYSTIC, we generate true color images of an Ocean planet, which corresponds to a geometry over the Pacific ocean, where we removed all continents, and of an Earth-like scenario, showing a configuration with Asia, Africa, Europe and the Indian Ocean in the scenery. In Figs. \ref{fig:images_ocean} and \ref{fig:images_earth} we show how the Ocean and Earth-like planets look at different phase angles ($\alpha$ = 0, 30, 60, 90 and 120\degr). These two configurations are the starting point to generate spatially unresolved spectra and phase curves of the Ocean and Earth-like configurations, averaging over the full disk of the simulations. We selected the Earth-like configuration to maximize the land surface component, and to let it cover almost completely the ocean glint feature at high phase angles ($\alpha$ > 90\degr), where the ocean glint presents a major effect. \\
\noindent In Figs. \ref{fig:images_ocean} and \ref{fig:images_earth} we also show the same configurations with realistic clouds simulated through the 3D CG approach. When showing spectra and phase curves for a cloudy planet, we always run the same configurations with 12 different cloud fields, one for each month of the year, randomly selecting the date. This will allow us to show the 1$\sigma$ range in the spectra and phase curves that represent the seasonal variability of the clouds.\\
\noindent In Tables \ref{tab:Ocean} and \ref{tab:Earth-like} we report the selected dates for the Ocean and Earth-like configurations, respectively. To better represent typical cloud properties over an ocean or land surfaces, we selected the cloud fields from the "ERA5 hourly product on pressure levels from 1940 to present" to be at UT 22:00 for the Ocean configuration, where the illuminated side of the planet maximized the Pacific ocean, and at UT 06:00, to have the illuminated side representing the Earth-like configurations described above. In Tables \ref{tab:Ocean} and \ref{tab:Earth-like} we report the calculated cloud properties of the different cloud configurations for $\alpha = 90$\degr, in particular their cloud cover, optical thickness, altitude and effective radius for liquid and ice water clouds. We also present the cumulative cloud cover for both liquid and ice water clouds as it depends non-trivially on their vertical overlap, while the cumulative optical thickness can be obtained as the sum of the individual ice and liquid water clouds optical thicknesses. We calculate the cloud properties only for the visible scene. The cloud cover is estimated using recursively Eq. \ref{eq:exp_ran_cloud_cover} to compute $c_{n+1/2}$ for each column and averaging over the visible scene. To estimate cloud altitudes more accurately in Tables \ref{tab:Ocean} and \ref{tab:Earth-like}, we use the procedure described in Appendix \ref{sec:appendix_A}, while MYSTIC simulations assume constant height levels over the full globe.\\
\noindent Tables \ref{tab:Ocean} and \ref{tab:Earth-like} show seasonal variability across different dates, particularly in cloud cover and optical thickness. They also reveal statistical differences in the average cloud properties between the Ocean and Earth-like configurations. Specifically, we find that cloud cover over the Ocean (cloud fields with UT 22:00) is about 5\% higher than over the Earth-like configuration. This is expected due to the presence of dry regions, such as the Sahara Desert, in the selected geometry. Additionally, the Earth-like configurations exhibit lower $\tau$ due to the lower cloud cover. We also observe that clouds tend to form at higher average altitudes over ocean surfaces and exhibit larger effective radii for both liquid and ice water clouds. These differences are in agreement with cloud patterns found in satellite observations and reanalysis products and allow us to represent in our spectra and phase curves of the Earth as an exoplanet the variability due to seasonal changes in cloud properties and their differences as a function of ocean versus land coverage.
\begin{table*}[h]
\caption{\label{tab:Ocean} Cloud properties for the 12 different cloud fields used for the Ocean planet scenario calculated for $\alpha$ = 90\degr. The last row shows the average properties for clouds over the Ocean planet.}
\centering
\resizebox{0.89\textwidth}{!}{
\begin{tabular}{c|c|cccc|cccc}
\toprule
& All clouds & \multicolumn{4}{c|}{Liquid water clouds} & \multicolumn{4}{c}{Ice water clouds} \\  
date & cc [\%] & cc [\%] & H [km] & $r_{\rm eff}$ [$\mu$m] & $\tau$ & cc [\%] & H [km] & $r_{\rm eff}$ [$\mu$m] & $\tau$ \\
\midrule
2023.01.15 UT22:00 & 58.1 & 47.3 & 1.83 & 9.3 & 7.14 & 51.1 & 4.51 & 50.7 & 0.54 \\
2023.02.16 UT22:00 & 59.6 & 48.5 & 1.79 & 9.5 & 7.22 & 53.8 & 4.97 & 50.9 & 0.66 \\
2023.03.29 UT22:00 & 64.6 & 54.2 & 1.56 & 9.2 & 7.00 & 59.9 & 4.36 & 48.1 & 0.66 \\
2023.04.10 UT22:00 & 57.7 & 49.5 & 1.67 & 9.4 & 6.62 & 52.2 & 5.43 & 50.1 & 0.61 \\
2023.05.17 UT22:00 & 63.1 & 56.6 & 1.67 & 9.4 & 6.88 & 59.5 & 4.45 & 51.0 & 0.69 \\
2023.06.21 UT22:00 & 64.7 & 57.7 & 1.54 & 9.4 & 9.46 & 59.5 & 5.14 & 49.2 & 0.62 \\
2023.07.14 UT22:00 & 66.6 & 59.0 & 1.56 & 9.8 & 9.55 & 57.0 & 6.18 & 48.8 & 0.69 \\
2023.08.06 UT22:00 & 62.6 & 55.4 & 1.59 & 9.7 & 9.92 & 57.5 & 4.66 & 48.5 & 0.64 \\
2023.09.12 UT22:00 & 61.7 & 53.6 & 1.71 & 9.7 & 7.86 & 57.8 & 4.96 & 49.4 & 0.69 \\
2023.10.07 UT22:00 & 61.2 & 54.1 & 1.53 & 9.3 & 7.45 & 55.4 & 4.28 & 48.7 & 0.61 \\
2023.11.27 UT22:00 & 63.7 & 51.7 & 1.64 & 9.0 & 5.61 & 59.5 & 4.23 & 50.7 & 0.65 \\
2023.12.30 UT22:00 & 61.9 & 49.0 & 1.61 & 9.0 & 6.30 & 54.7 & 4.43 & 50.0 & 0.68 \\
\midrule
average & 62.1 & 53.1 & 1.64 & 9.4 & 7.58 & 56.5 & 4.80 & 49.7 & 0.64 \\
\bottomrule
\end{tabular}
}
\end{table*}
\begin{table*}[h]
\caption{\label{tab:Earth-like} Cloud properties for the 12 different cloud fields used for the Earth-like planet scenario calculated for $\alpha$ = 90\degr. The last row shows the average properties for clouds over the Earth-like planet.}
\centering
\resizebox{0.89\textwidth}{!}{
\begin{tabular}{c|c|cccc|cccc}
\toprule
& All clouds & \multicolumn{4}{c|}{Liquid water clouds} & \multicolumn{4}{c}{Ice water clouds} \\   
date & cc [\%] & cc [\%] & H [km] & $r_{\rm eff}$ [$\mu$m] & $\tau$ & cc [\%] & H [km] & $r_{\rm eff}$ [$\mu$m] & $\tau$ \\
\midrule
2023.01.22 UT06:00 & 50.5 & 38.1 & 1.58 & 8.7 & 5.11 & 45.5 & 4.95 & 47.5 & 0.45 \\
2023.02.08 UT06:00 & 57.1 & 41.5 & 1.46 & 8.5 & 5.81 & 52.5 & 3.91 & 46.5 & 0.55 \\
2023.03.02 UT06:00 & 53.2 & 38.3 & 1.49 & 8.6 & 5.23 & 47.7 & 3.44 & 45.6 & 0.53 \\
2023.04.09 UT06:00 & 50.1 & 35.0 & 1.48 & 8.8 & 4.51 & 44.2 & 3.77 & 46.1 & 0.45 \\
2023.05.14 UT06:00 & 61.5 & 36.5 & 1.52 & 8.7 & 5.62 & 57.2 & 3.89 & 42.0 & 0.68 \\
2023.06.05 UT06:00 & 63.0 & 37.1 & 1.64 & 8.3 & 5.56 & 57.5 & 3.69 & 42.7 & 1.10 \\
2023.07.26 UT06:00 & 61.4 & 36.2 & 1.79 & 8.6 & 5.19 & 54.7 & 3.59 & 42.1 & 0.70 \\
2023.08.29 UT06:00 & 55.9 & 35.9 & 1.54 & 8.7 & 5.61 & 48.3 & 4.19 & 42.6 & 0.56 \\
2023.09.18 UT06:00 & 60.2 & 40.0 & 1.51 & 8.5 & 5.36 & 54.4 & 3.85 & 42.2 & 0.64 \\
2023.10.05 UT06:00 & 57.5 & 35.0 & 1.53 & 8.5 & 5.42 & 51.8 & 3.24 & 42.0 & 0.58 \\
2023.11.12 UT06:00 & 60.7 & 41.5 & 1.50 & 8.7 & 6.29 & 56.3 & 3.71 & 45.1 & 0.70 \\
2023.12.24 UT06:00 & 52.9 & 41.5 & 1.51 & 8.5 & 5.53 & 48.4 & 4.35 & 47.4 & 0.50 \\
\midrule
average & 57.0 & 38.0 & 1.55 & 8.6 & 5.44 & 51.5 & 3.88 & 44.3 & 0.62 \\
\bottomrule
\end{tabular}
}
\end{table*}

\subsection{Impact of the 3D Cloud Generator}
\label{sec:3DCG_results}

\begin{figure*}[h]
    \centering
    \includegraphics[width=1\linewidth]{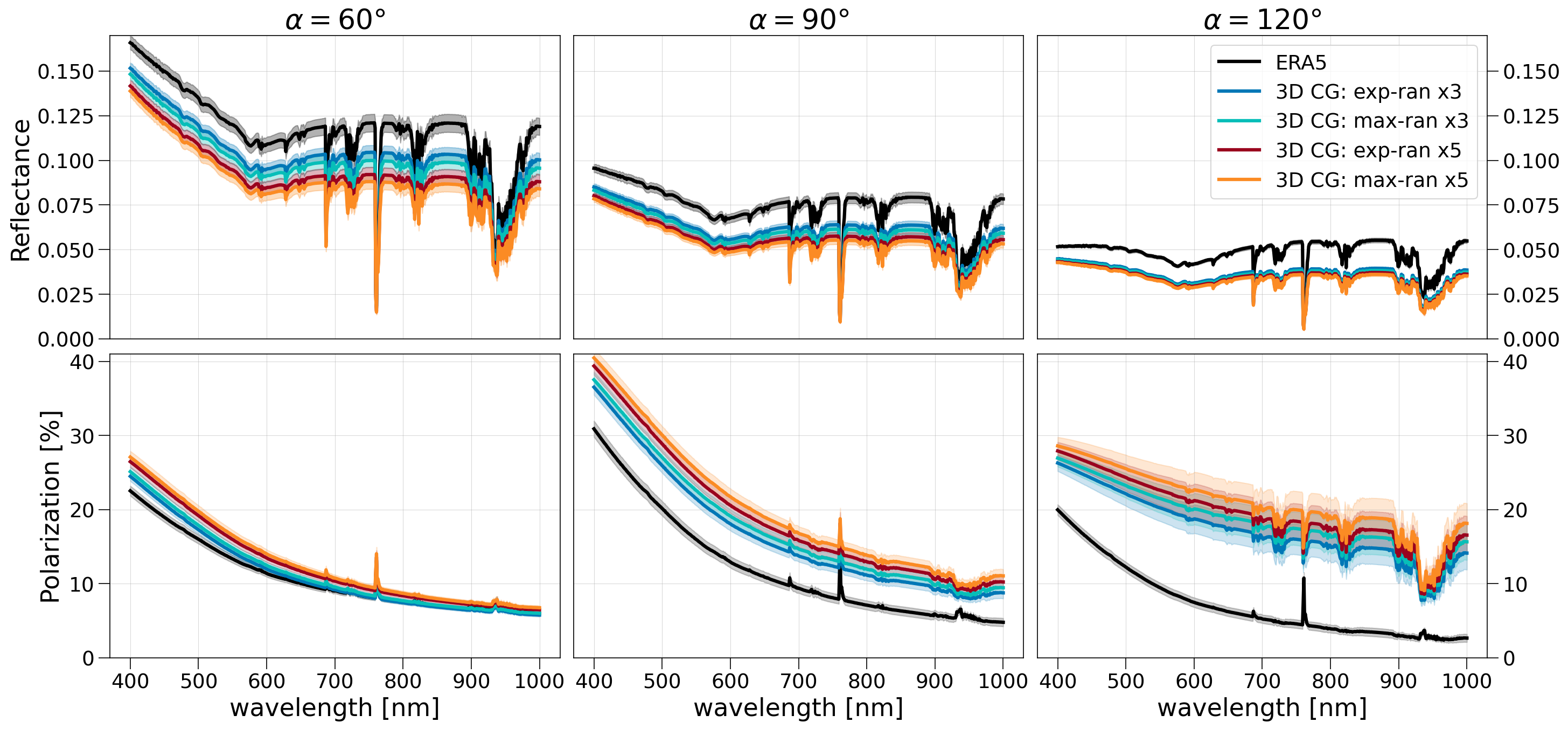}
    \caption{Reflected light (first row) and polarized light (second row) spectra showing the influence of the 3D CG approach applied to the ERA5 reanalysis data for the clouds, as compared to ERA5 data themselves (black line). The 3D CG is ran assuming different vertical overlap schemes: EXP-RAN and MAX-RAN and different zoom-in factors (x3 and x5). Different columns refer to spectra at different phase angles $\alpha$: 60, 90, 120\degr.}
    \label{fig:spectra_CG}
\end{figure*}

To investigate the impact of cloud representation, particularly the cloud radiative response in the 3D radiative transfer simulations, we selected an ocean surface with constant wind speed of 10\,m\,s$^{-1}$ as our configuration (Fig. \ref{fig:images_ocean}). Starting from cloud properties from the ERA5 reanalysis product, we introduce sub-grid cloud inhomogeneities using the 3D CG approach as described in Sec. \ref{sec:3DCG}. \\
\noindent In Appendix \ref{sec:appendix_B}, we analyze the impact of the zoom-in factor on the radiative transfer effects of clouds and find that it quickly converges to a stationary value, even at a zoom factor of 3. In Fig. \ref{fig:spectra_CG}, we show the reflectance and polarization spectra of an Ocean planet with different cloud models for three different phase angles: 60, 90 and 120\degr. We can clearly see that the reflectance of the ERA5 cloud configurations is greatly overestimated when not accounting for the sub-grid variability of the clouds. This strongly influences the albedo of the planet, in particular at high phase angles. For the polarization spectra, we find that the 3D CG representation has an impact on the slope of the spectra at small phase angles ($\alpha$ = 60\degr), it shows an important difference in the polarization at $\alpha$ = 90\degr, while the effect becomes substantial at high phase angles ($\alpha$ = 120\degr). In particular, this large discrepancy at high phase angle can be explained by the effect of the ocean glint. Making the clouds more "patchy" and increasing their horizontal inhomogeneities allows for the glint to be less masked by the clouds and to greatly increase the polarization compared to the ERA5 clouds. Additionally, the 1$\sigma$ spread due to cloud variability increases at larger phase angles. This is because the ocean glint feature has its strongest impact on polarization spectra and phase curves at these angles. Running simulations at large phase angles with different cloud fields, each containing a different spatial cloud distribution, results in varying degrees of cloud coverage over the ocean glint region. As a result, some cases exhibit a more pronounced glint effect, while others are more obscured by clouds, leading to an increased spread in the simulations.\\
\noindent In polarization, we also observe a change in the behavior of atmospheric absorption lines. We define an absorption feature as a case where the absorption line in polarization falls below the continuum, whereas we refer to it as emission when the line appears above the continuum. In particular, the O$_2$-A line around 770~nm is shown in emission for most of the polarized spectra, while it changes behaviour between the 3D CG simulations (absorption) and the ERA5 clouds (emission) at $\alpha$ = 120\degr. This change in behaviour is due again to the different cloud properties, in particular cloud optical depth and cloud height, and how they are represented in the radiative transfer simulations. A different behaviour is also observed for the water band around 950~nm, where the feature appears in absorption only when including sub-grid cloud variability, while otherwise shown with weak emission features. These last finding is in agreement with \cite{trees2022}, where they describe how sensitive polarization is, contrary to intensity spectra, to the ocean glint feature. In their simulations, water bands appear in absorption in polarization when there is an ocean glint feature not fully covered by clouds, while they appear in emissions in the case of full cloud cover over the ocean glint or for a dry surface. With our simulations we show that, even when applying realistic cloud cover and horizontal cloud patterns and taking into account the sub-grid variability, the ocean glint feature still retains its typical effect of changing from emission to absorption depending on the surface type of the planet. This difference will become even more significant later for the comparison with the Earth-like scenario, where the ocean glint will be hidden by the African continent for $\alpha$ > 90\degr. \\
\begin{figure*}[h]
    \centering
    \includegraphics[width=1\linewidth]{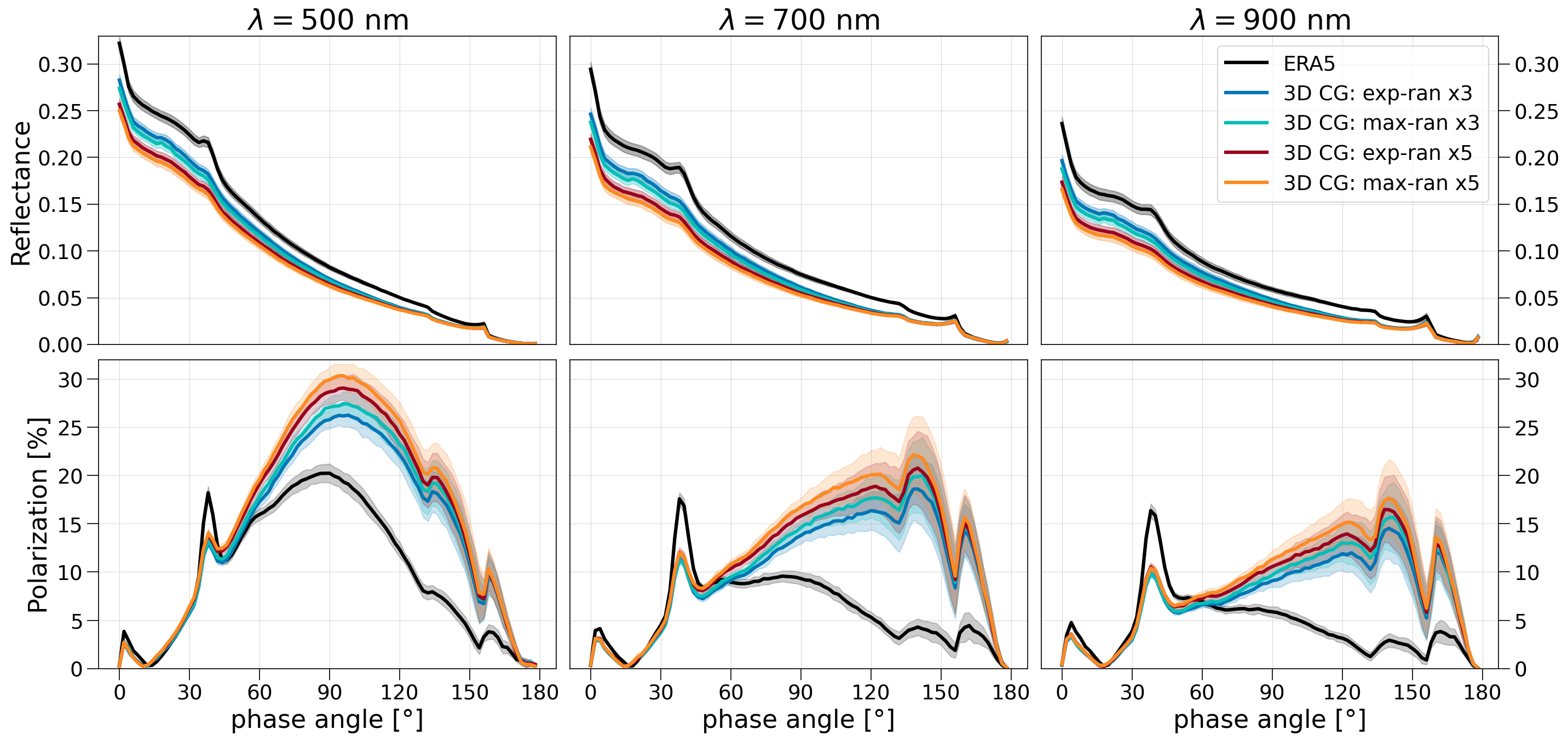}
    \caption{Reflected light (first row) and polarized light (second row) phase curves showing the influence of the 3D CG approach applied to the ERA5 reanalysis data for the clouds, as compared to ERA5 data themselves (black line). The 3D CG is ran assuming different vertical overlap schemes: EXP-RAN and MAX-RAN and different zoom-in factors (x3 and x5). Different columns refer to different wavelengths ($\lambda$): 500, 700, 900~nm.}
    \label{fig:phase_CG}
\end{figure*}
\noindent In Fig. \ref{fig:spectra_CG} we not only highlight the impact of the 3D CG, but we also study the effect of assuming different schemes for the vertical overlap of clouds, nominally the MAX-RAN and EXP-RAN overlaps. As described in Sec. \ref{sec:3DCG}, the EXP-RAN overlap generally increases the total cloud cover of a single cloudy pixel at the TOA, making the clouds more reflective but also more realistic by introducing some random overlap even for continuous cloudy layers. This is also the case for the disk-integrated simulations, where the simulations with EXP-RAN overlap have a slightly increased reflectance compared to the same MAX-RAN simulations with the 3D CG. Polarization exhibits the opposite behavior. The EXP-RAN overlap scheme leads to more cloud cover, blocking a larger part of the ocean, where glint reflection produces polarization. It also obscures more of the lower atmosphere, which contributes to polarization through Rayleigh scattering. As a result, the overall polarization in the spectra is lower than in the MAX-RAN overlap simulations.\\
\noindent We also explore the effect of different zoom-in factors for each ERA5 pixel in the 3D CG. Comparing between a zoom-in factor of x3 and x5 per size (meaning that we are creating 9 and 25 sub-pixels, respectively), we study the impact of the increase in horizontal resolution on the reflectance and polarized spectra of spatially unresolved planets. Intuitively, the more pixels we have in a simulation, the better we can resolve the 3D cloud structure and the closer to reality it should look like. However, running global simulations with pixels size that can reach up to the fractal dimension of clouds, where cloud structures can be resolved, is unfeasible. Starting from the ERA5 reanalysis product, which has a pixel size of around 31~km, with a x3 and x5 zoom-in we obtain pixel sizes of approximately 9 and 6~km, respectively (or 0.08\degr and 0.05\degr), still far from the fractal dimension of clouds, yet reaching a great level of details and accuracy. Comparing the x3 and x5 spectra with the same vertical overlap scheme (MAX-RAN or EXP-RAN), we see that by increasing the zoom-in factor we also slightly decrease the reflectance of the planet and slightly increase its polarization, as already expected by our convergence study in Fig. \ref{fig:convergence}. However, a zoom-in factor of 5 is significantly more computationally expensive than a zoom-in factor of 3, while not substantially affecting the results. Additionally, we clearly see that the impact of the 3D CG applied to the ERA5 clouds is much more significant than the differences among different zoom-in factors and vertical overlap schemes. Thus, from now on we will be using the 3D CG with EXP-RAN overlap and x3 zoom-in factor as the ground-truth model to treat clouds in our simulations.\\
\noindent For the reflected light spectra, we find that reflectance decreases with increasing phase angles, as expected due to the progressively smaller illuminated portion of the planet. For the polarized spectra, we show that the largest polarization fraction can be found around $\alpha$ = 90\degr, where there is a peak in the polarization due to Rayleigh scattering. Also the slope of the polarized spectra is affected by both Rayleigh scattering and the ocean surface. The slope is particularly affected at $\alpha$ = 120\degr due to the ocean glint polarization, which is spectrally independent and dominates over Rayleigh scattering at large phase angles, especially when there are more gaps in the clouds introduced by the 3D CG.\\
\noindent After assessing the influence of the 3D CG on the reflectance and polarized spectra, in Fig. \ref{fig:phase_CG} we also assess its impact on the phase curves, again for the Ocean planet configuration. As discussed for the reflectance spectra, the 3D CG has an impact on the reflectance of the planet, lowering it at all wavelengths ($\lambda$ = 500, 700 and 900~nm.) But the impact is even larger when comparing the ERA5 clouds with the 3D CG approach for the phase curves in polarization, where it significantly increases the amount of linear polarization expected for a disk-integrated observation of the planet. This is a consequence of introducing sub-grid cloud variability, thus allowing radiation to travel into the 3D cloud inhomogeneous structure, instead of photons being reflected by smeared homogeneous clouds as represented in ERA5. The effect of the different vertical overlap schemes (MAX-RAN and EXP-RAN) and of the zoom-in parameter is the same as described for the spectra, and it is still less significant than the difference between including or not sub-grid variability and inhomogeneities through the 3D CG. \\
\noindent As expected, for $\lambda$ = 500~nm we find the largest reflectance and polarization due to Rayleigh scattering, while it decreases moving towards 700~nm and 900~nm. However, the impact of the 3D CG appears increasingly relevant after $\alpha$ > 60\degr, where the impact of the ocean glint feature becomes larger and the sub-grid cloud inhomogeneity impacts the radiative transfer. This happens at all wavelengths, confirming the weak chromatic effect of the ocean glint feature. At small phase angles ($\alpha \sim 5^\circ$), we find the glory feature, which is linked to the optical properties of the cloud droplets. This feature does not seem to be affected by the 3D CG approach. Around $\alpha$ = 40\degr, we find the cloudbow feature, which is a consequence of the first internal reflection of the liquid water droplets. Its location and height is strongly linked with the microphysical properties of clouds, in particular their optical thickness, effective radius and particle composition. As shown, the cloudbow feature appears more pronounced in the polarization spectra, allowing to break down possible retrieval degeneracies about cloud properties as discussed in \cite{sterzik2020}. We observe that the cloudbow feature is significantly influenced by the 3D CG in both reflected light and polarized phase curves. This effect can be attributed to the higher in-cloud $LWC$ in the 3D CG configurations compared to ERA5, which leads to a smaller peak in polarization. In the phase curves, we observe that, for $\lambda = 500~nm$, the polarization peak caused by Raileigh scattering occurs at approximately $\alpha$ = 90\degr for the ERA5 clouds, whereas it shifts to larger phase angles when using the 3D CG. For larger wavelenght, Raileigh scattering weakens. Additional features found at $\alpha$ = 138\degr and 158\degr arise from the scattering properties of ice clouds and depend on their optical characteristics \citep{emde2017}.

\subsection{Clouds zoom-out}
\label{sec:zoom_out_results}

\begin{figure*}[h]
    \centering
    \includegraphics[width=1\linewidth]{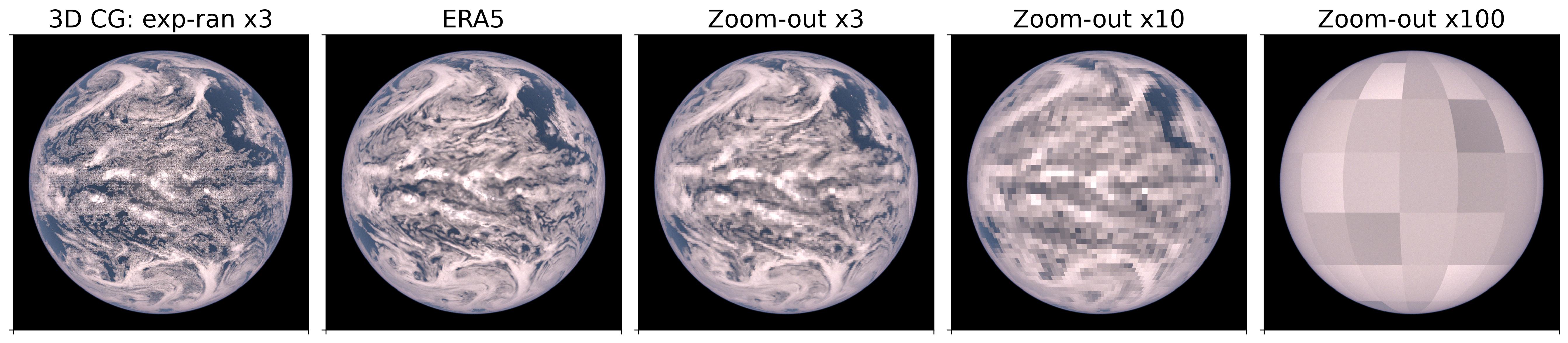}
    \caption{True Color images of the Ocean planet scenario showing the impact of different grid resolutions in representing the clouds. We show the 3D CG image with a zoom-in x3 factor compared to the ERA5 image, and zoomed-out images with factors x3, x10 and x100. Reducing the grid size we notice an increase in the total reflectance of the planet.}
    \label{fig:images_zoom_out}
\end{figure*}

After assessing the impact of the 3D CG, we want to study the effect of averaging out cloud properties on larger and larger grid scales in order to mimic exoplanet atmosphere simulations. Using different zoom-out factors, from x3 ($\sim$ 84~km, 0.75\degr), x10 ($\sim$ 278~km, 2.5\degr), x100 ($\sim$ 2670~km, 24\degr) and a single pixel simulation, we average out layer by layer the cloud properties, conserving the optical thickness as described in Sec. \ref{sec:zoom-out}. In Fig. \ref{fig:images_zoom_out}, we show the Ocean planet images for the different cloud models, from the 3D CG (EXP-RAN overlap, zoom-in x3) to the ERA5 image and the different zoom-out cases for $\alpha$ = 0\degr. Reducing the amount of pixels, we notice an increase in reflectance of the planet, as the $LWC$ and $IWC$ present in the sub-grid need to be smeared out in larger pixels. This effect is already present in the comparison between the 3D CG and the ERA5 images, as discussed above.\\
\begin{figure*}[h]
    \centering
    \includegraphics[width=1\linewidth]{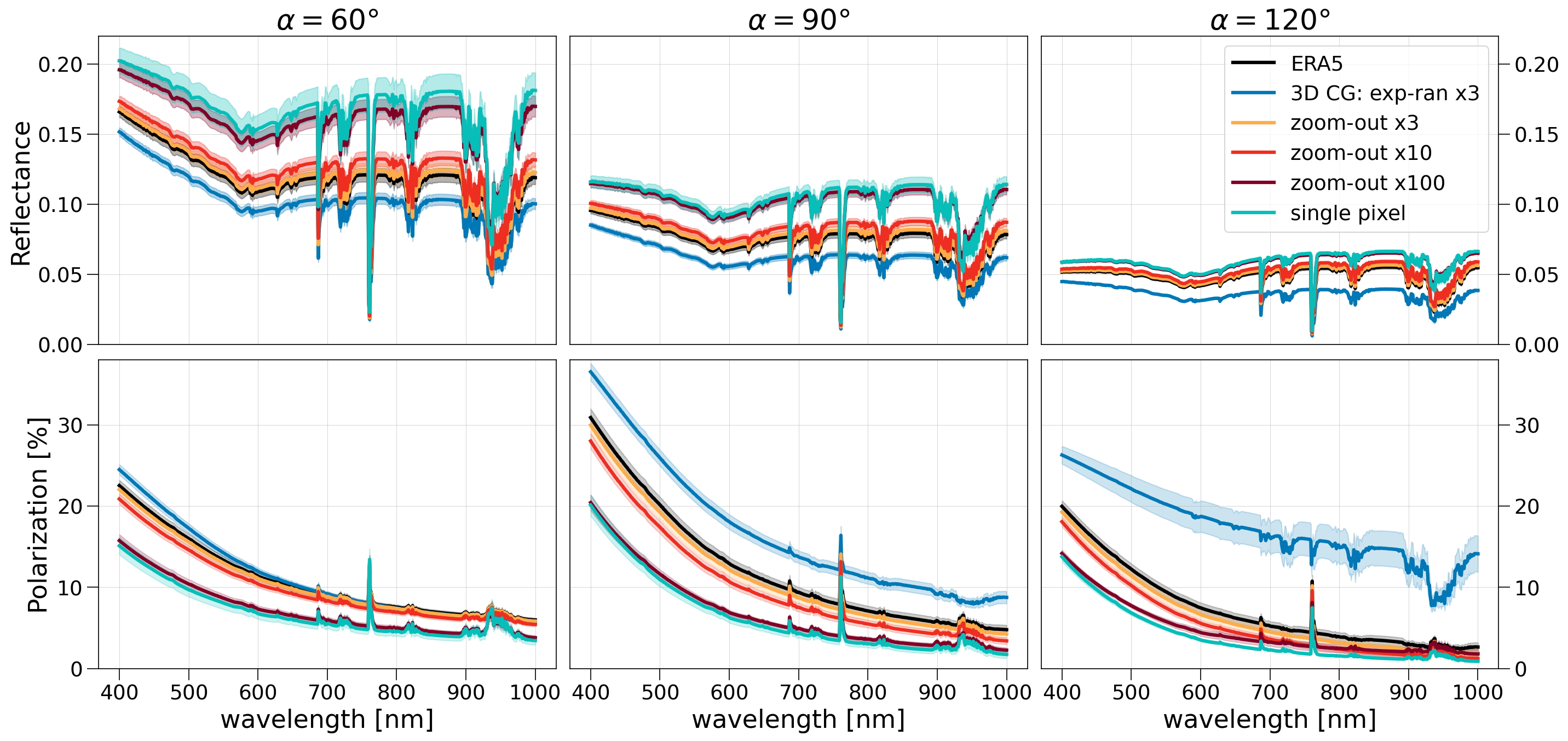}
    \caption{Reflected light (first row) and polarized light (second row) spectra showing the influence of the number of pixels in the simulations. From the 3D CG EXP-RAN zoom-in x3 and ERA5 simulations, we apply the cloud zoom-out algorithm, with zoom-out factors x3, x10, x100, until a single-pixel simulation. With the zoom-out, the reflectance of the planet gets substantially overestimated, while the polarization is influenced both in the spectral slope and molecular lines. Different columns refer to spectra at different phase angles ($\alpha$): 60, 90, 120\degr.}
    \label{fig:spectra_zoom_out}
\end{figure*}
\begin{figure*}[h]
    \centering
    \includegraphics[width=1\linewidth]{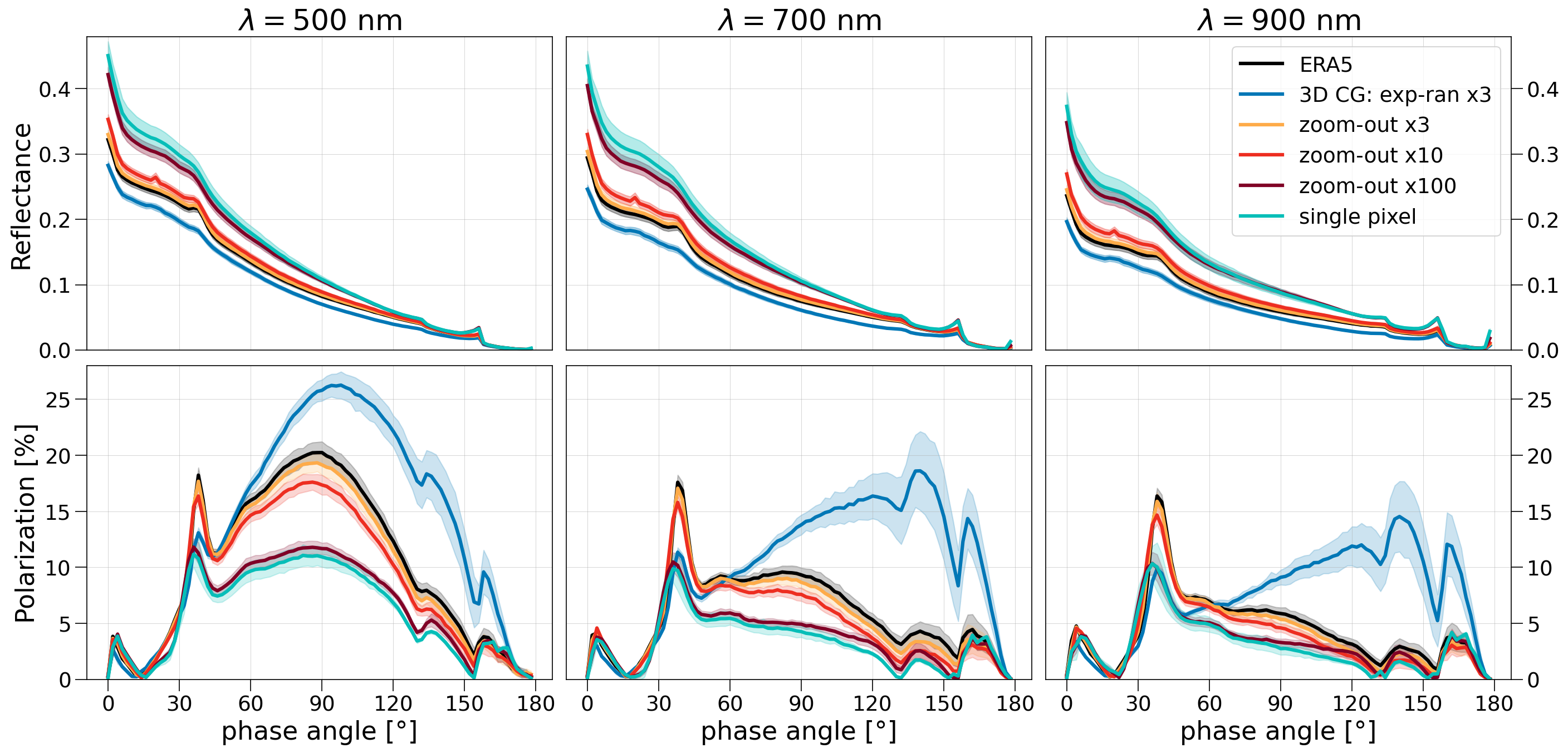}
    \caption{Reflected light (first row) and polarized light (second row) phase curves showing the influence of the cloud zoom-out algorithm. Different columns refer to different wavelengths ($\lambda$): 500, 700, 900~nm.}
    \label{fig:phase_zoom_out}
\end{figure*}
\noindent In Fig. \ref{fig:spectra_zoom_out}, we show the impact on the reflected and polarized light spectra of running simulations with coarser grids. As a reference, we have the 3D CG and the ERA5 cloud models from the previous plot, and new spectra with zoom-out factors of x3, x10, x100, until arriving to a single pixel simulation. As expected, the general trend we observe is an increase of the reflectance of the planet at all phase angles, without any large impact on the spectral slope. While the zoom-out ×3 and zoom-out ×10 simulations remain comparable with ERA5, the zoom-out ×100 and single-pixel simulations show a significant increase in reflectance. This occurs because they can no longer accurately represent the 3D cloud structure of the planet, instead displaying nearly homogeneous cloud properties across the entire disk. For the polarization, we observe again a large jump between the zoom-out x10 and zoom-out x100 simulations, also with a substantial impact on the slope of the polarized spectra. Again, also the O$_2$A band and water bands are affected by the coarser cloud grid. \\
\noindent Studying the effect of the zoom-out on the phase curves (Fig. \ref{fig:phase_zoom_out}), we observe a large impact from the zoom-out x10 to the zoom-out x100 case, as the planet becomes almost homogeneous. The impact is still affecting both the reflected and polarized light spectra, with a larger effect in polarization for $\alpha > 90^\circ$. In addition, the cloudbow feature ($\alpha \sim 40^\circ$) is impacted by the zoom-out factor in polarization, with the zoom-out x100 being almost identical to the single pixel simulations. \\
\noindent Comparing the zoom-in and zoom-out studies, we can clearly see that the zoom-in process, even with just a factor of x3, has a much more important effect on both spectra and phase curves if compared with a zoom-out of x3. This suggests the importance of treating the sub-grid cloud variability and finding better solutions on how to parametrize clouds in exoplanet atmosphere simulations, where such high resolution will not be possible. However, it remains fundamental to treat clouds in a more realistic way, as their radiative effect has a strong influence on both spectra and phase curves, in particular for polarization studies. This suggests how polarization spectra contain more information about cloud properties and might be used to improve degeneracies among different models and interpretations for exoplanet studies.\\
\noindent Additionally, in Appendix \ref{sec:appendix_C}, we conduct sensitivity studies on the impact of the cloud droplet effective radius and find no significant differences when the optical thickness of the clouds is conserved.

\subsection{Hyperspectral albedo maps}
\label{sec:albedo_results}

\begin{figure*}
    \centering
    \includegraphics[width=1\linewidth]{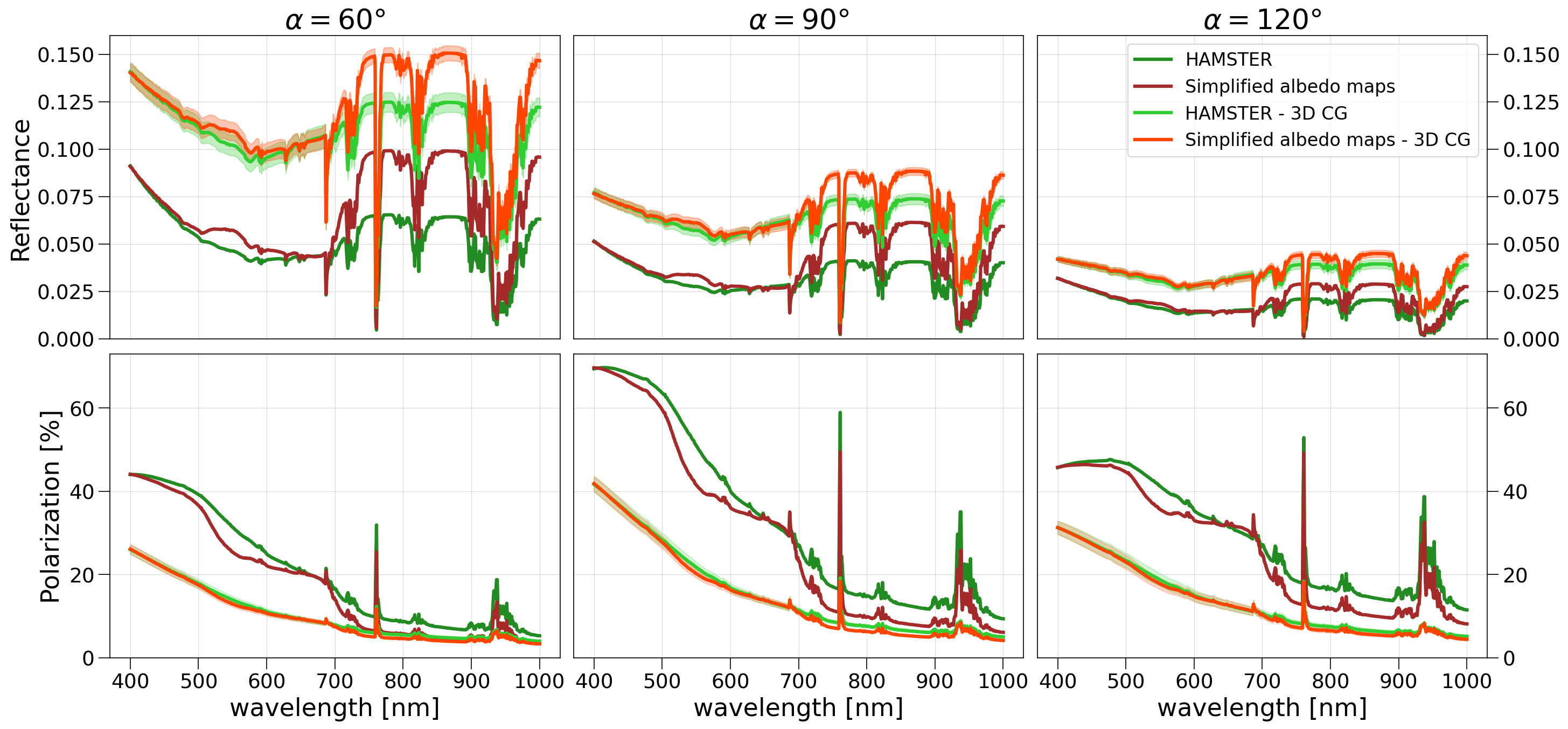}
    \caption{Reflected light (first row) and polarized light (second row) spectra comparing HAMSTER with simplified hyperspectral albedo maps, generated using a linear combination of five ECOSTRESS spectra. Different columns refer to spectra at different phase angles ($\alpha$): 60, 90, 120\degr.}
    \label{fig:HAMSTER_comparison_spectra}
\end{figure*}

\begin{figure*}
    \centering
    \includegraphics[width=1\linewidth]{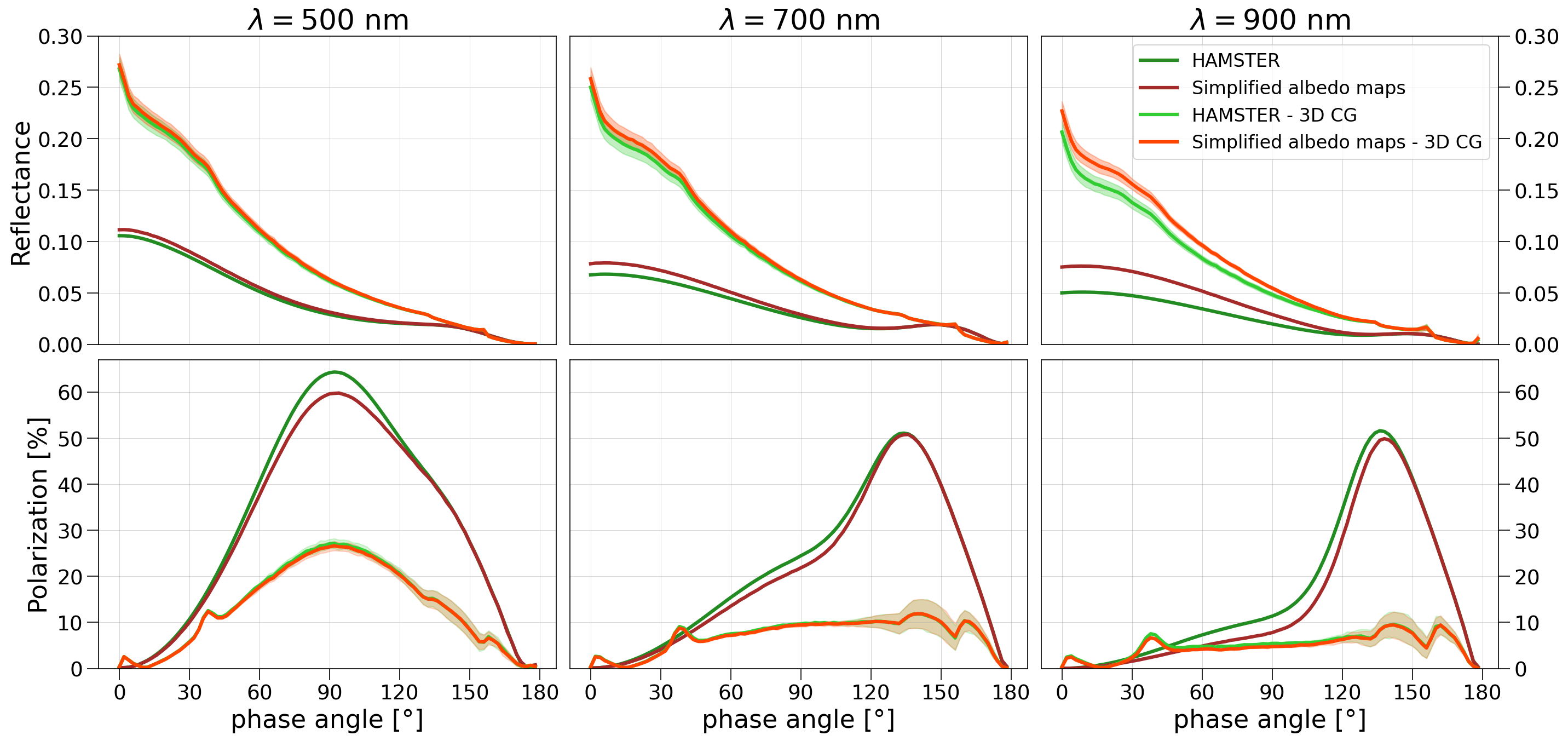}
    \caption{Reflected light (first row) and polarized light (second row) phase curves comparing HAMSTER with simplified hyperspectral albedo maps, generated using a linear combination of five ECOSTRESS spectra. Different columns refer to different wavelengths ($\lambda$): 500, 700, 900~nm.}
    \label{fig:HAMSTER_comparison_phase}
\end{figure*}

To study the sensitivity of surface properties, we first investigate the role of homogeneous and inhomogeneous surface wind speed maps on the ocean glint feature in Appendix \ref{sec:appendix_D} and find no significant differences within the 1$\sigma$ cloud variability spread.\\
\noindent To simulate the Earth-like planet scenario (Fig. \ref{fig:images_earth}), we incorporate the modeling of land surfaces. Using the hyperspectral albedo maps dataset HAMSTER \citep{Roccetti2024}, we include a wavelength-dependent treatment of surface reflectance for land surfaces. This is crucial for accurately representing the planet's albedo and capturing spectral features resulting from surface reflection. We also note that previous works \citep{gordon2023, Kofman2024} that included wavelength-dependent surface albedo treatments typically coupled spectral libraries with land surface types from satellite observations to define the albedo in each simulation pixel. This approach may lead to highly biased surface reflectance spectra, as spectral libraries do not account for the fact that forests consist of a combination of components, such as leaves and soils, not just the spectrum of a single leaf. By using HAMSTER, we are able to incorporate the first remote sensing-calibrated dataset of wavelength-dependent surface albedo maps, enhancing our representation of land surfaces. In particular, we find a substantially reduced VRE feature (see Fig. \ref{fig:VRE}), which has been highly overestimated in previous Earth-like simulations.\\
\noindent In Fig. \ref{fig:HAMSTER_comparison_spectra}, we show the difference between HAMSTER and the simplified hyperspectral albedo maps. The VRE feature and the green bump (around 550~nm) are clearly visible in the reflectance spectra. However, when using a linear combination of ECOSTRESS spectra from MODIS land surface type maps, these features are significantly overestimated in reflected light spectra. This overestimation is particularly pronounced at small phase angles, nearly doubling the expected continuum beyond the VRE feature (750 nm) for a cloudless planet. In cloudy scenarios, where we apply the 3D CG treatment, the effect remains substantial, highlighting the impact of HAMSTER’s improved surface albedo model. In particular, with HAMSTER, the green bump is completely washed out in the disk-integrated spectra, even in a clear atmosphere.\\
\noindent In polarized spectra, we observe the opposite trend. Since all land surfaces are treated as Lambertian, they do not polarize radiation and instead scale inversely with intensity. As a result, the VRE and green bump features are underestimated in polarized spectra for a cloudless planet. The difference is most pronounced at $\alpha$ = 90\degr in cloudless simulations, reaching approximately a 25\% difference in polarization, but is less significant for cloudy planets, where the difference falls within the 1$\sigma$ cloud variability spread.\\
\noindent In Fig. \ref{fig:HAMSTER_comparison_phase}, we also examine the effect on phase curves. At $\lambda$ = 500 and 700~nm, we find no significant differences between HAMSTER and the simplified hyperspectral albedo maps. However, at $\lambda$ = 900~nm, within the VRE peak, the simplified albedo maps produce an unphysical increase in reflectance. This effect is particularly noticeable in cloudless simulations, where the reflectance is increased by 50\% when using the simplified albedo maps, and remains significant even for a cloudy planet, showing an increase of approximately 20\% at small phase angles. In polarization, the impact on phase curves is only relevant for the cloudless scenario at wavelengths beyond the VRE. We also see a difference in the phase curve shape for polarized light if compared to the Ocean planet scenario (Fig. \ref{fig:phase_wind}). Introducing different albedo surfaces, we see not only the peak of polarization around $\alpha$ = 90\degr due to Rayleigh scattering, but also another jump, becoming evident for $\lambda$ = 700 and 900~nm. This additional jump is due to the reflectance of different albedo components compared to the ocean surface, such as vegetation or deserts, and they show their typical signatures at longer wavelengths. This shows that polarization is a more powerful diagnostic tool for surface features than intensity alone when looking at disk-integrated spectra and phase curves.\\
\noindent Additionally, we also address the impact of the albedo seasonal variability in HAMSTER in the spectra and phase curves. The results are shown in Appendix \ref{sec:appendix_E}.

\subsection{Comparing the Ocean and Earth-like planets}
\label{sec:comparison}

\begin{figure*}[h]
    \centering
    \includegraphics[width=1\linewidth]{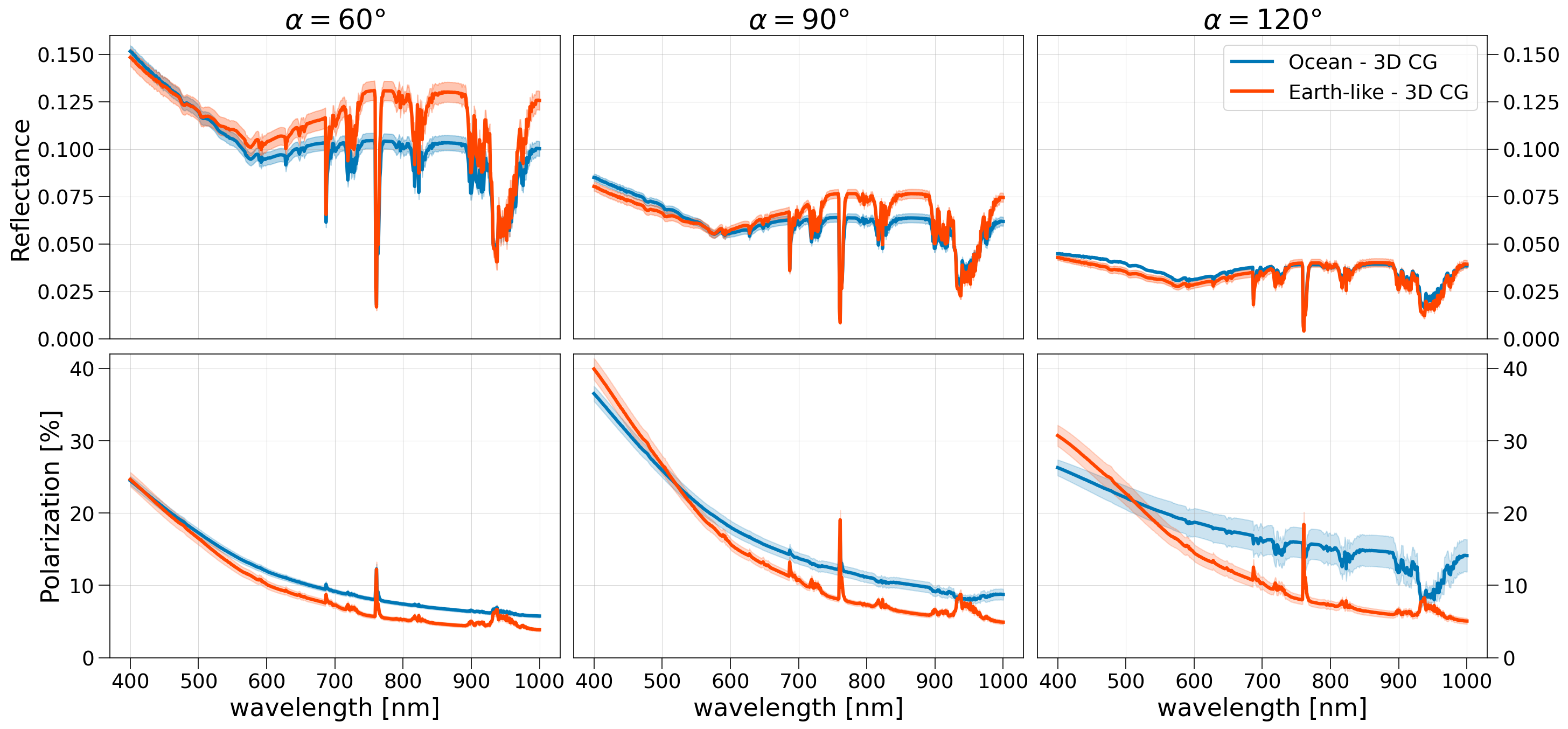}
    \caption{Comparison among spectra in reflected (first row) and polarized light (second row) of the Ocean and Earth-like planet scenarios. Different columns refer to different phase angles ($\alpha$): 60, 90, 120\degr.}
    \label{fig:spectra_comparison}
\end{figure*}

After introducing various surface and cloud modeling improvements, we now want to compare between the ground truth Ocean and Earth-like planet scenarios. We are comparing the following cases:
\begin{itemize}
    \item Ocean surface (including BPDF) with the 3D CG clouds and their 1$\sigma$ spread;
    \item Earth-like scenario including ocean surface treated with the BPDF (but ocean glint almost always covered by land) and hyperspectral albedo maps with the 3D CG clouds and their 1$\sigma$ spread.
\end{itemize}
In Fig. \ref{fig:spectra_comparison} we show the reflected and polarized light differences in the spectra due to the different models. We notice a significant spread between the Ocean and Earth-like scenario, way beyond the cloud variability in the models found in the 1$\sigma$ spread (shaded areas). This is particularly evident for $\alpha$ = 60\degr in intensity, while the difference gets larger for $\alpha$ = 120\degr in polarization, where we observe a different behaviour both in the spectral sloples of the models, their continuum in the near-infrared (NIR) and in the spectral lines behaviour. Due to the presence (Ocean scenario) and absence (Earth-like scenario) of the ocean glint feature, we see a different behaviour of the water bands around 950~nm, as they are shown in absorption for the Ocean planet and in emission for the Earth-like case. This effect is already present in the $\alpha$ = 90\degr case, but gets enhanced at larger phase angles. Additionally, at large phase angles, cloud variability is significantly larger for the Ocean planet than for the Earth-like scenario. This effect is most pronounced at larger phase angles, where the spread is dominated by ocean glint. \\
\begin{figure*}[h]
    \centering
    \includegraphics[width=1\linewidth]{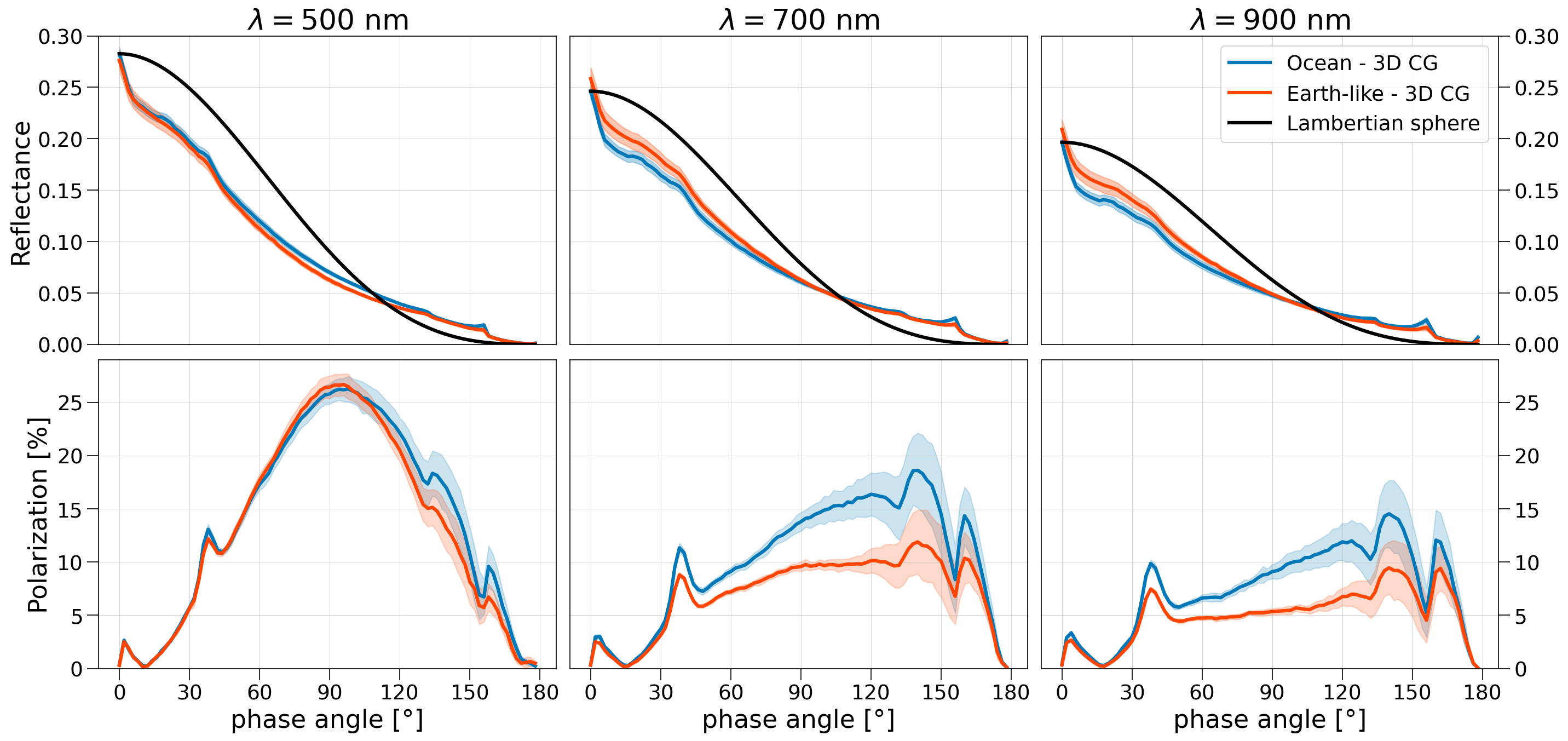}
    \caption{Reflected light (first row) and polarized light (second row) phase curves comparing the Ocean and Earth-like planet scenarios. Different columns refer to different wavelengths ($\lambda$): 500, 700, 900~nm.}
    \label{fig:phase_comparison}
\end{figure*}
\noindent In Fig. \ref{fig:phase_comparison}, we study the same differences in reflected and polarized light over the phase curves again at three different wavelengths: $\lambda$ = 500, 700 and 900~nm. Among the Ocean and Earth-like scenario, large differences are found only in the polarization case, in particular for $\lambda$ = 700 and 900~nm. We note an increase in polarization due to the ocean glint which goes beyond the cloud variability spread we introduce in the models. To distinguish among any possible features in reflected light is more challenging than in polarization. For comparison, we also include the phase curve of a purely Lambertian sphere, modeled using the Lambertian phase function
\begin{equation}
g(\alpha)_L = \frac{\sin(\alpha) + (\pi - \alpha)\cos(\alpha)}{\pi}.    
\label{eq:lamb_phase}
\end{equation}
The Lambertian phase function is scaled by the geometric albedo of the Ocean planet case at each wavelength.\\
\noindent In Appendix \ref{sec:appendix_F}, we also present the same spectral comparison for the Ocean and Earth-like planet scenarios, but in terms of albedo instead of reflectance (Fig. \ref{fig:albedo_comparison}).

\section{Discussion and conclusions}
\label{sec:conclusions}

In this paper, we assess the importance of incorporating inhomogeneous surfaces and atmospheres into 3D radiative transfer models of Earth-like exoplanets. Specifically, we study the effect of implementing, for the first time, wavelength-dependent surface albedo maps derived from Earth remote sensing observations. In parallel, we address the importance of integrating inhomogeneous 3D cloud maps into our models, with a focus on sub-grid cloud variability and cloud inhomogeneities. While exoplanet forward models must be computationally efficient to explore key parameters and perform spectral retrievals for interpreting observations, this study focuses on assessing the impact of different model complexities on the resulting spectra and phase curves of an Earth-like exoplanet observed in reflected and polarized light.
Future studies should further assess the importance of fine details in modeling observations for the next generation of telescopes, particularly in relation to the typical integration times required to achieve a sufficient signal-to-noise ratio.\\
\noindent By studying the effects of the 3D CG and the cloud "zoom-out" process, we conclude that cloud grid size in simulations plays a critical role in the reflectance and polarized spectra and phase curves. Specifically, without accounting for sub-grid cloud variability and cloud inhomogeneities introduced by the 3D CG, the planet’s reflectance is significantly overestimated. Additionally, clouds become too widespread within grid cells, which lowers the planet’s linear polarization. The effect of insufficient spatial resolution implies that too simple models may be misleading towards expectations of a too high reflectance, and too low polarization. This is particularly relevant in exoplanet models, where homogeneous cloud properties are often assumed, potentially leading to biases in data interpretation. Although obtaining such fine-scale details for exoplanets is impossible and developing models that resolve cloud properties at these detailed spatial and vertical scales remains infeasible, improved parametrization schemes must be developed to preserve the radiative effects of clouds in coarser-resolution models. Our results should inform the appropriate resolution at which radiative transfer codes, coupled with General Circulation Models (GCMs), should be run to avoid biasing the interpretation of observed spectra and phase curves. Furthermore, in our sensitivity analysis regarding the effective radius of cloud particles, we find that its impact is minimal when the optical thickness of grid cells is conserved. This highlights the need to appropriately scale cloud optical thickness during the "zoom-out" process for coarser models. \\
\noindent All spectra and phase curves generated for cloudy planets are accompanied by a 1$\sigma$ cloud variability spread, which helps in building ground-truth models of Earth as an exoplanet. This spread also aids in determining whether potential diagnostic features used to distinguish exoplanet properties might be obscured by cloud variability.\\
\noindent We introduce the use of a new dataset to account for the pixel-by-pixel spectral variation of surface albedo, as well as its seasonal variability. Using HAMSTER \citep{Roccetti2024}, we demonstrate that previous attempts to incorporate wavelength-dependent surface reflectance for different surface types, such as forests and deserts, greatly overestimated reflectance. For instance, the VRE is often overestimated by nearly 100\% due to the assumption that forested areas could be represented by laboratory spectra of individual leaves. This study underscores the necessity of incorporating realistic spectral albedo maps to accurately capture spectral features in the reflectance and polarized spectra of spatially unresolved exoplanets. As shown by \cite{Gomez-Barrientos2023}, a wavelength-dependent surface albedo model more accurately retrieves reflected light spectra than a uniform albedo model, even in the presence of clouds. Here, we demonstrate that constructing such albedo maps requires careful consideration. Relying solely on laboratory-based wavelength-dependent measurements can lead to a substantial overestimation of the VRE feature in reflected light spectra, even with realistic cloud coverage. However, detecting seasonal variability, such as snow cover changes, remains extremely challenging, as these effects are masked by the planet’s cloud abundance and variability. Additionally, we investigate the role of surface wind speed in shaping ocean glint features but find no significant impact on the spectra or phase curves.\\
\noindent After constructing these advanced models, we compare their effects on reflected and polarized light spectra and phase curves.
These results indicate that polarization is far more sensitive to surface features, as evidenced by differences between Ocean and Earth-like planet scenarios in reflected and polarized light. In polarization, spectral slopes and absorption lines differ more prominently than in intensity, particularly at large phase angles. Despite the improved cloud modeling, we still observe unique water line features in polarization, as reported by \cite{trees2022} for homogeneous planets. Specifically, water lines appear in absorption when ocean glint is present but in emission for dry planets. These findings suggest that combining polarization with intensity-only spectroscopy can greatly enhance the characterization of rocky exoplanets and provide greater diagnostic power to differentiate surface and atmospheric properties, reducing retrieval degeneracies.\\
\noindent Future studies should explore the feasibility of conducting polarized light observations with upcoming telescopes and missions, assessing whether sufficient contrast can be achieved for further exoplanet characterization. Polarization is particularly advantageous because stellar contamination can be significantly reduced, as most FGK-type stars are nearly unpolarized. Contrast estimates for observing rocky exoplanets in reflected and polarized light will be presented in the second paper of this series, using the fully improved modeling setup described in this work.\\
\noindent The improved modeling approaches presented here will be validated against a large catalog of Earthshine observations obtained in polarized light to assess the model’s ability to reproduce observed spectral features. This validation will be addressed in the third paper of this series.\\
\noindent In conclusion, the ground-truth spectra and phase curves simulated for Earth-like and Ocean planet scenarios will be made available to the exoplanet community. These datasets can improve predictions for next-generation telescopes and instruments, validate other exoplanet models, and facilitate studies of Earth as an exoplanet. 

\section{Data availability}
All spectra and phase curve data are openly accessible via a Jupyter notebook on the GitHub repository \url{https://github.com/giulia-roccetti/Earth_as_an_exoplanet_Part_I}. Additionally, we also provide public access to the 3D Cloud Generator algorithm at \url{https://github.com/giulia-roccetti/3D_Cloud_Generator}.

\begin{acknowledgements}
We thank the anonymous referee for the useful and constructive comments, which greatly helped improving this paper. GR and JVS were supported by the Munich Institute for Astro-, Particle and BioPhysics (MIAPbP) which is funded by the Deutsche Forschungsgemeinschaft (DFG, German Research Foundation) under Germany´s Excellence Strategy – EXC-2094 – 390783311.
\end{acknowledgements}

\bibliographystyle{aa} 
\bibliography{bibliography} 

\begin{appendix}

\section{ERA5 inhomogeneous pressure level heights}
\label{sec:appendix_A}

To compute the heights of the ERA5 pressure levels, we start from assuming hydrostatic equilibrium and an ideal gas law for dry air:
\begin{equation}
    \frac{dP(z)}{dz} = - \rho(z) g = -\frac{P(z)g}{R_L T(z)},
\label{eq:dpdz}
\end{equation}
where $P(z)$ is the pressure as a function of height, $R_L\simeq$ 287.05\,J\,kg$^{-1}$\,K$^{-1}$ is the specific gas constant for dry air and $g \simeq$ 9.807\,m\,s$^{-2}$ is Earth's gravitational acceleration. We also assume temperature to be linearly dependent on height, within every layer:
\begin{equation}
    T(z) = T_i + (z-z_i)\frac{T_{i+1}-T_i}{z_{i+1}-z_i}.
\end{equation}
Integrating Eq. \ref{eq:dpdz} between level $i$ and $i+1$ we obtain the following equation for the thickness of the layer:
\begin{equation}
    \Delta z_i = \ln{\left(\frac{P_{i}}{P_{i+1}}\right)} \cdot \frac{R_L(T_{i+1}-T_i)}{g \ln{\left(\frac{T_{i+1}}{T_i}\right)}}.
\end{equation}
The height of a given layer $i$ is then computed by summing all the layers below:
\begin{equation}
    z_i = \sum_{k=0}^i \Delta z_k.
\end{equation}

\section{Convergence of the 3D Cloud Generator}
\label{sec:appendix_B}

\begin{figure*}[h]
    \centering
    \includegraphics[width=1.0\linewidth]{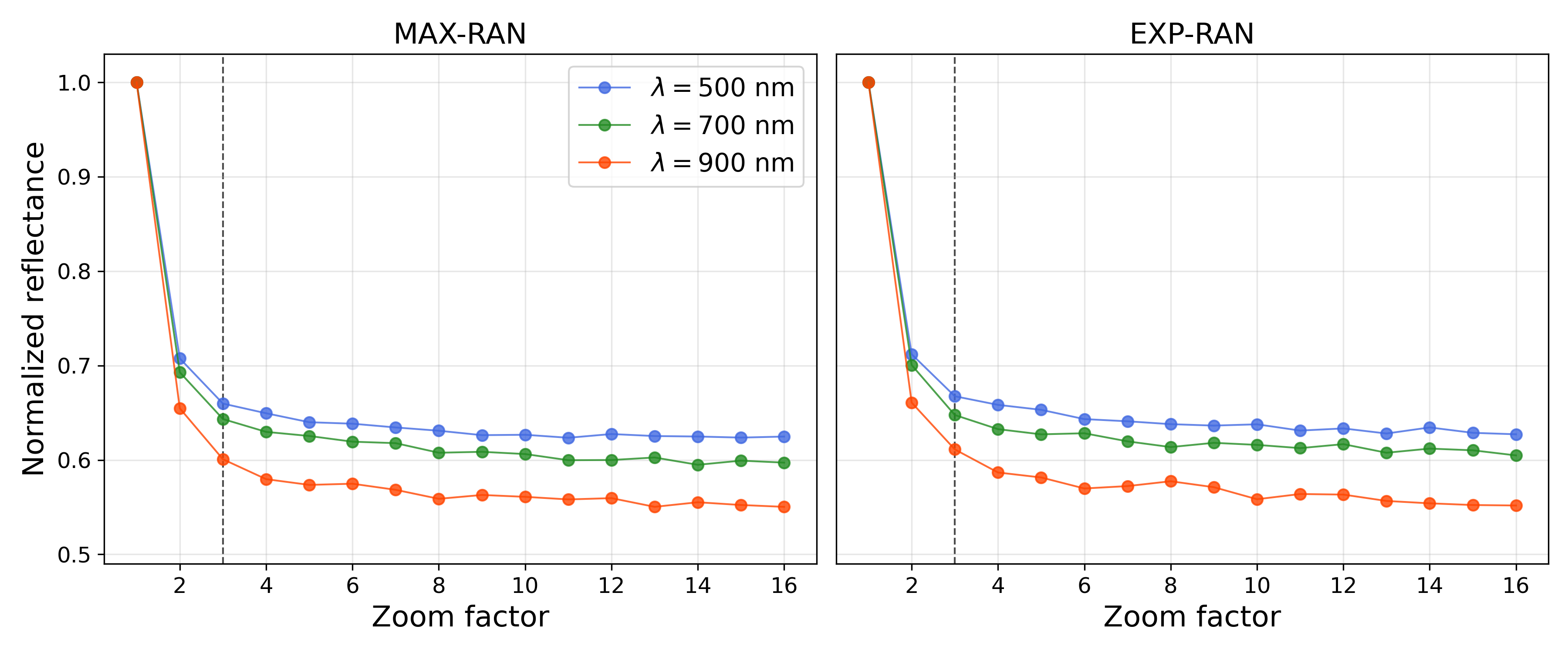}
    \caption{Convergence study on the impact of the zoom-in factor on the radiative response of clouds in a region spanning 10\degr in both latitude and longitude. Both the MAX-RAN and the EXP-RAN overlap methods are shown, for three wavelengths in the visible and NIR range. The reflectance is normalized to its value without zoom-in. The vertical dashed line represents a zoom-in factor of 3, which will be used for most of the results below, as it provides the optimal balance between convergence and computational cost.}
    \label{fig:convergence}
\end{figure*}

The 3D CG described in Sec. \ref{sec:3DCG} can generate finer cloud distributions starting from ERA5 realanysis data for an arbitrary zoom-in factor. In this Appendix, we analyze the impact of the zoom-in factor on the radiative transfer effects of clouds. Specifically, we generate cloud fields with zoom-in factors ranging from 2 to 16 of a region spanning 10\degr in both latitude and longitude, using a constant surface albedo of 0.3. This is done because running a global simulation with such a fine cloud grid up to a zoom factor of 16 is too computationally expensive. A sensor is placed at the center of this patch, at an altitude of 1000~km, with a 30\degr aperture, ensuring that the entire patch remains within its field of view. The results are shown in Fig. \ref{fig:convergence} for both the MAX-RAN and EXP-RAN overlap schemes and for the 500, 700 and 900~nm wavelengths. As shown in the figure, introducing a finer grid of clouds greatly impacts the normalized reflectance in radiative transfer calculations, leading to a decrease of approximately 40\%, depending on the wavelength. Additionally, the radiative response to the 3D CG quickly converges to a stationary value already for a zoom factor of $3$ for all wavelengths and for both methods. Since this provides the best balance between convergence and computational cost, we adopt a zoom-in factor of 3 for most of the results presented in this paper.

\section{Impact of the cloud droplet effective radius}
\label{sec:appendix_C}

\begin{figure*}[h!]
    \centering
    \includegraphics[width=1\linewidth]{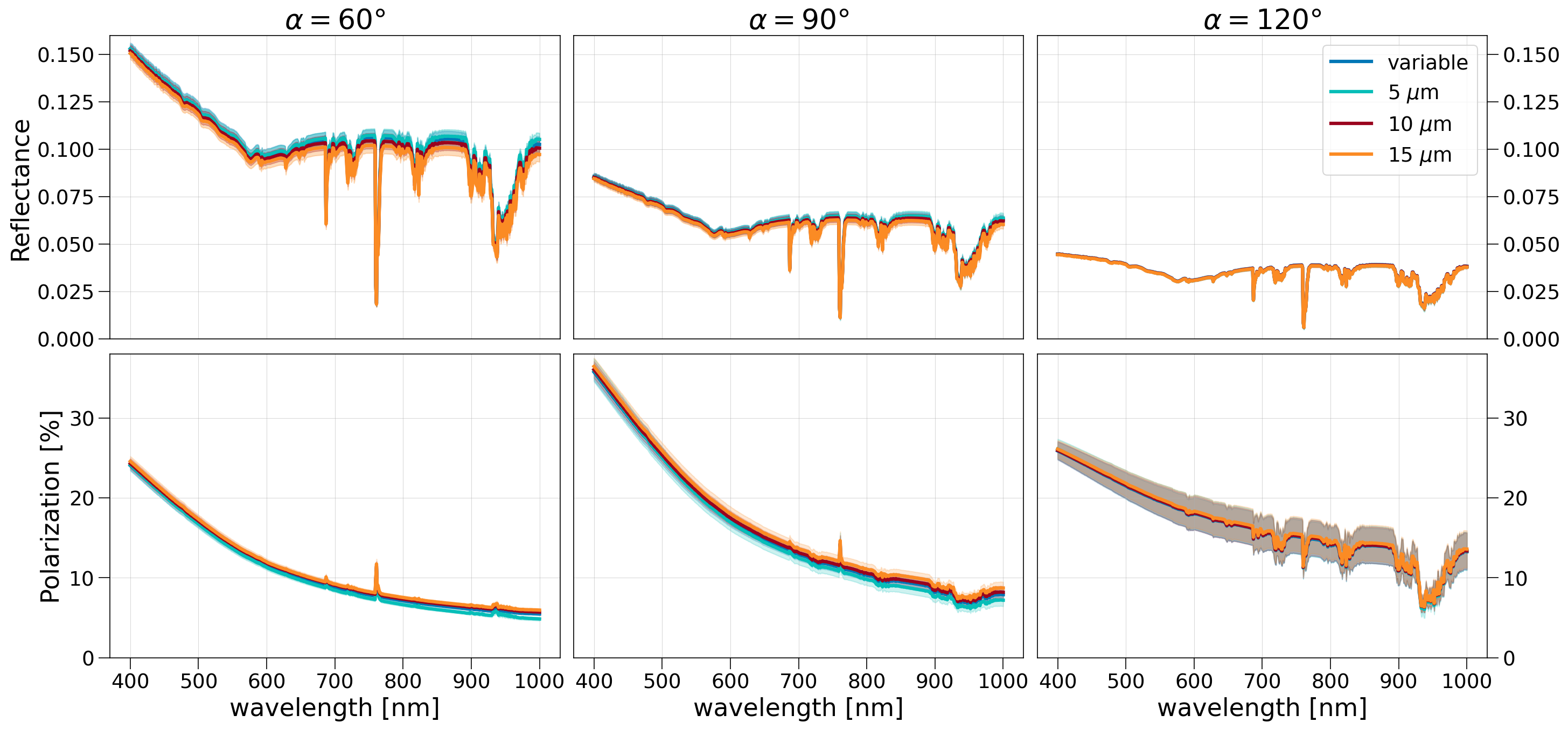}
    \caption{Reflected light (first row) and polarized light (second row) spectra showing the influence of the effective radius of cloud particles. Here, we compare a variable effective radius from the ECMWF parametrization to constant effective radius values, but always conserving the optical thickness of each gridbox. Different columns refer to spectra at different phase angles ($\alpha$): 60, 90, 120\degr.}
    \label{fig:spectra_eff_radius}
\end{figure*}

\begin{figure*}[h!]
    \centering
    \includegraphics[width=1\linewidth]{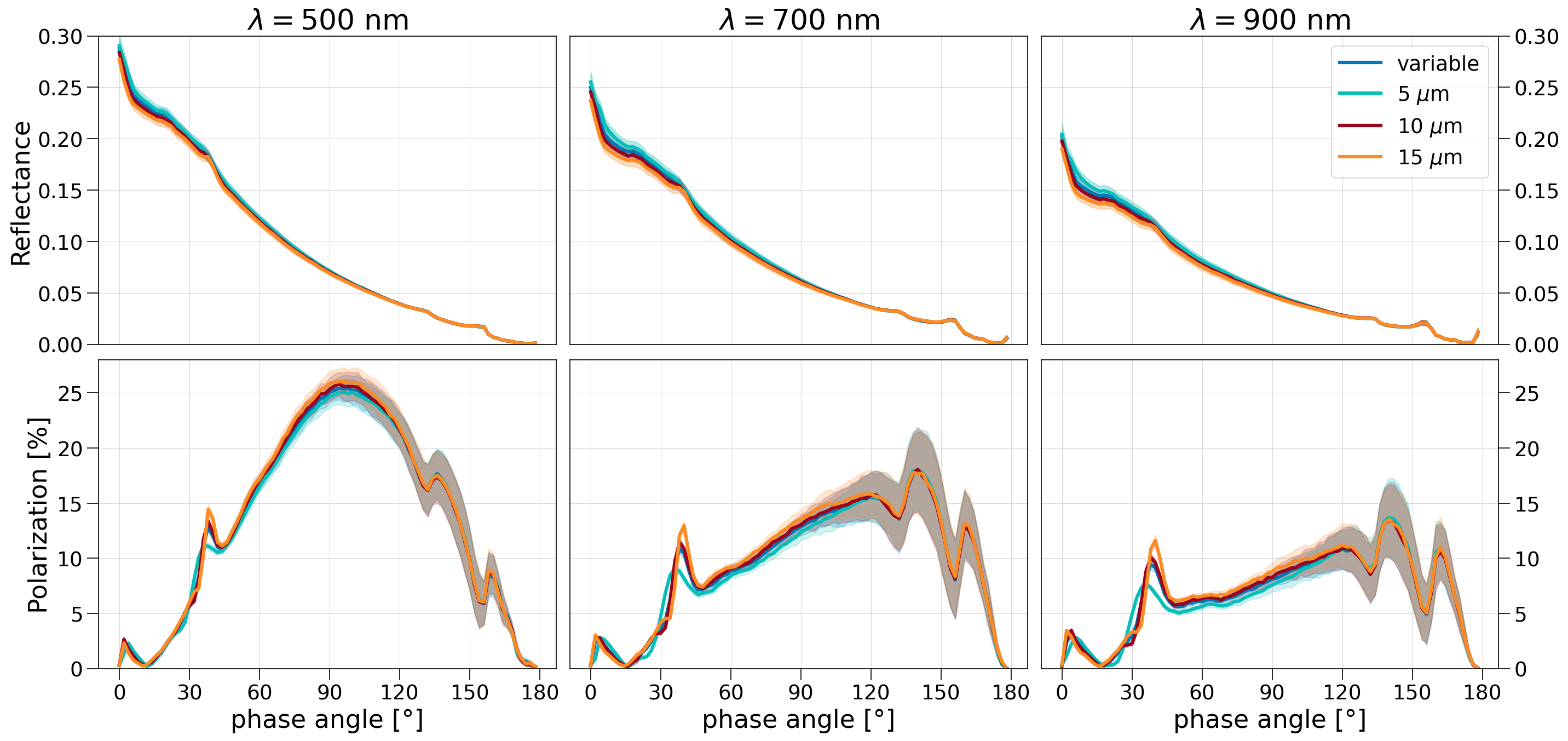}
    \caption{Reflected light (first row) and polarized light (second row) phase curves showing the influence of a constant effective radius, while conserving the optical thickess of the gridbox. Different columns refer to different wavelengths ($\lambda$): 500, 700, 900~nm.}
    \label{fig:phase_eff_radius}
\end{figure*}

As an additional parameter, we test the role of the effective radius of liquid and ice water droplets on the spectra and phase curves over an Ocean planet. Starting from the 3D CG, we conserve the optical thickness of clouds scaling the $LWC$ and $IWC$ while changing the effective radius to constant values of 5, 10 and 15~$\mu$m. We then compare it with the variable effective radius distribution obtained with the ERA5 parametrization. When conserving the optical thickness, we did not find any significant variations for all spectra in both reflected and polarized light (Fig. \ref{fig:spectra_eff_radius}). This suggests that only introducing variations in the optical thickness of clouds has a direct impact on the spectra, and since the effective radius enters in the calculation of $\tau$, it plays a role only when the $LWC$ and $IWC$ are not adjusted accordingly to conserve $\tau$. However, in the phase curves (Fig. \ref{fig:phase_eff_radius}) we observe the impact of the effective radius on the cloudbow feature ($\alpha \sim 40^\circ$), in both reflectance and polarization, greatly enhanced in the polarized phase curves. In polarization, we notice both a shift towards larger phase angles and an increase in the cloudbow feature for larger effective radii, showing again the sensitivity of linear polarization in assessing cloud particle size through the cloudbow feature. This effect gets enhanced at longer wavelengths ($\lambda$ = 900~nm). All the other differences are within the 1$\sigma$ cloud variability and cannot be clearly distinguished.


\section{Impact of the wind speed}
\label{sec:appendix_D}

\begin{figure*}[h]
    \centering
    \includegraphics[width=1\linewidth]{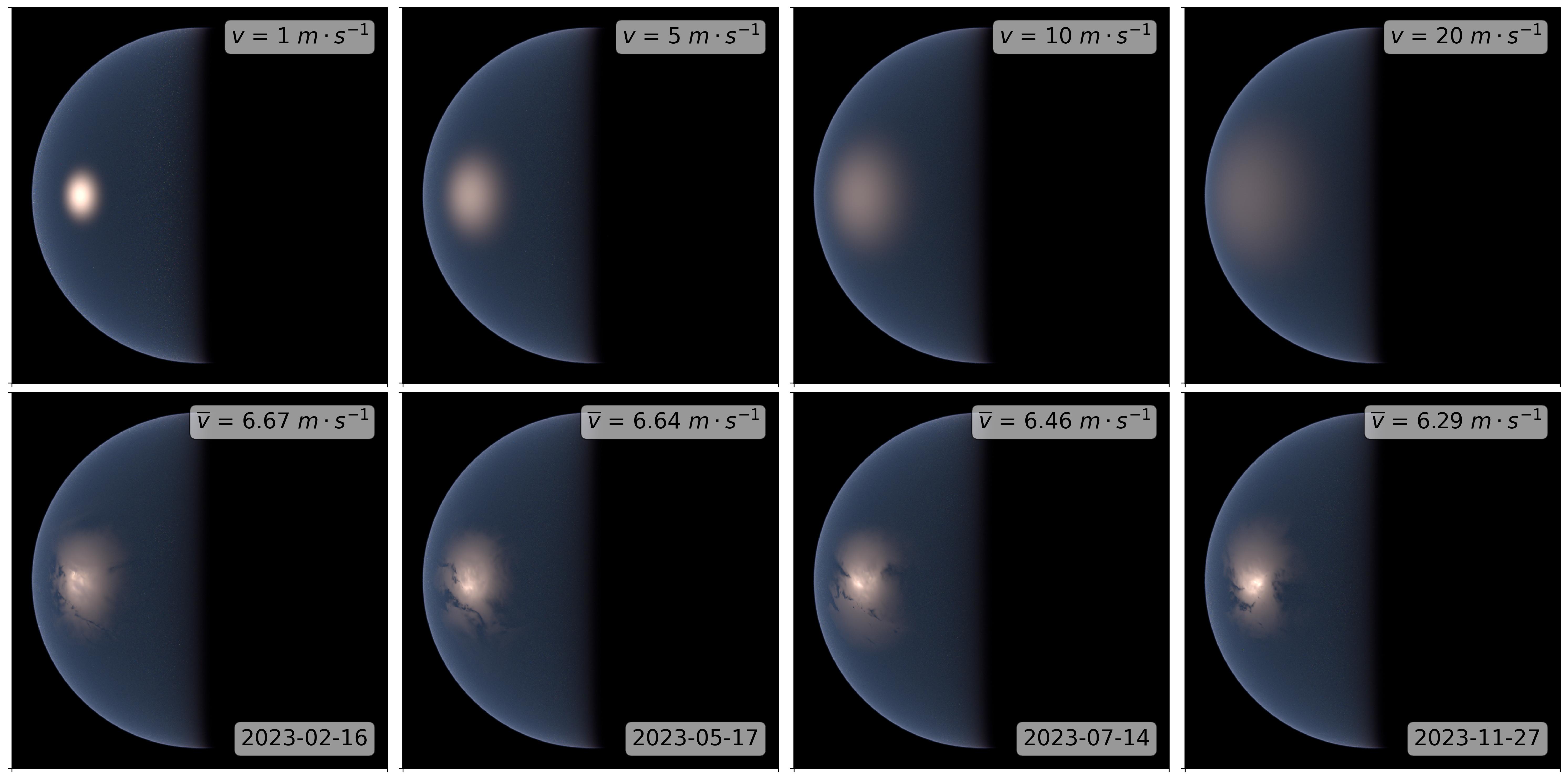}
    \caption{True colour image of a cloud-free Ocean planet at $\alpha$ = 90\degr with different surface wind speed maps. In the first row, we use a constant wind speed and notice the impact on the brightness and size of the ocean glint. In the second row, we use realistic wind speed maps from ERA5, and we notice the inhomogenous shape of the ocean glint. We report the average wind speed from the various ERA5 fields in the sub-figures.}
    \label{fig:images_wind}
\end{figure*}

\begin{figure*}[h]
    \centering
    \includegraphics[width=1\linewidth]{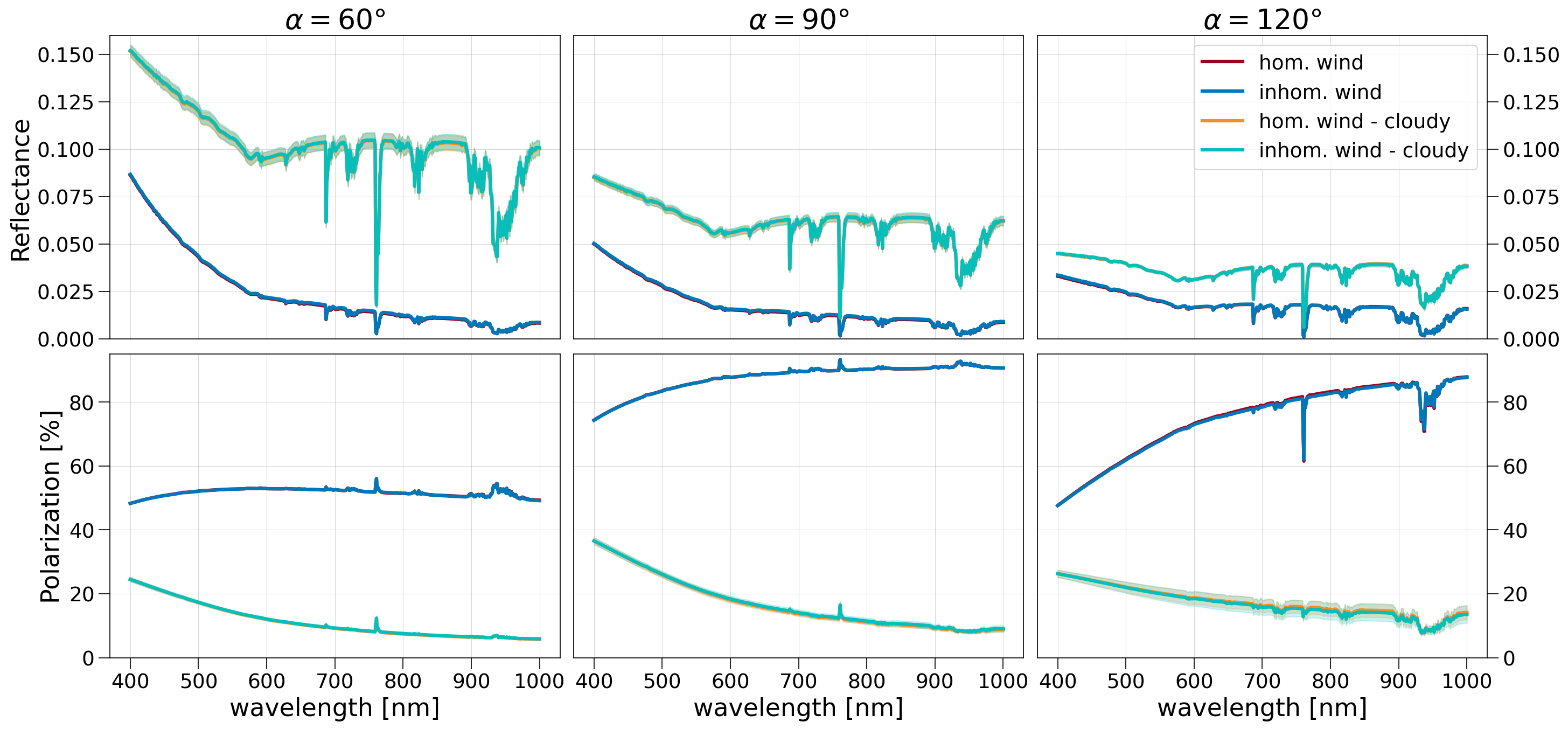}
    \caption{Reflected light (first row) and polarized light (second row) spectra showing the influence of homogeneous and inhomogeneous wind speed maps. Different columns refer to spectra at different phase angles ($\alpha$): 60, 90, 120\degr.}
    \label{fig:spectra_wind}
\end{figure*}

\begin{figure*}[h]
    \centering
    \includegraphics[width=1\linewidth]{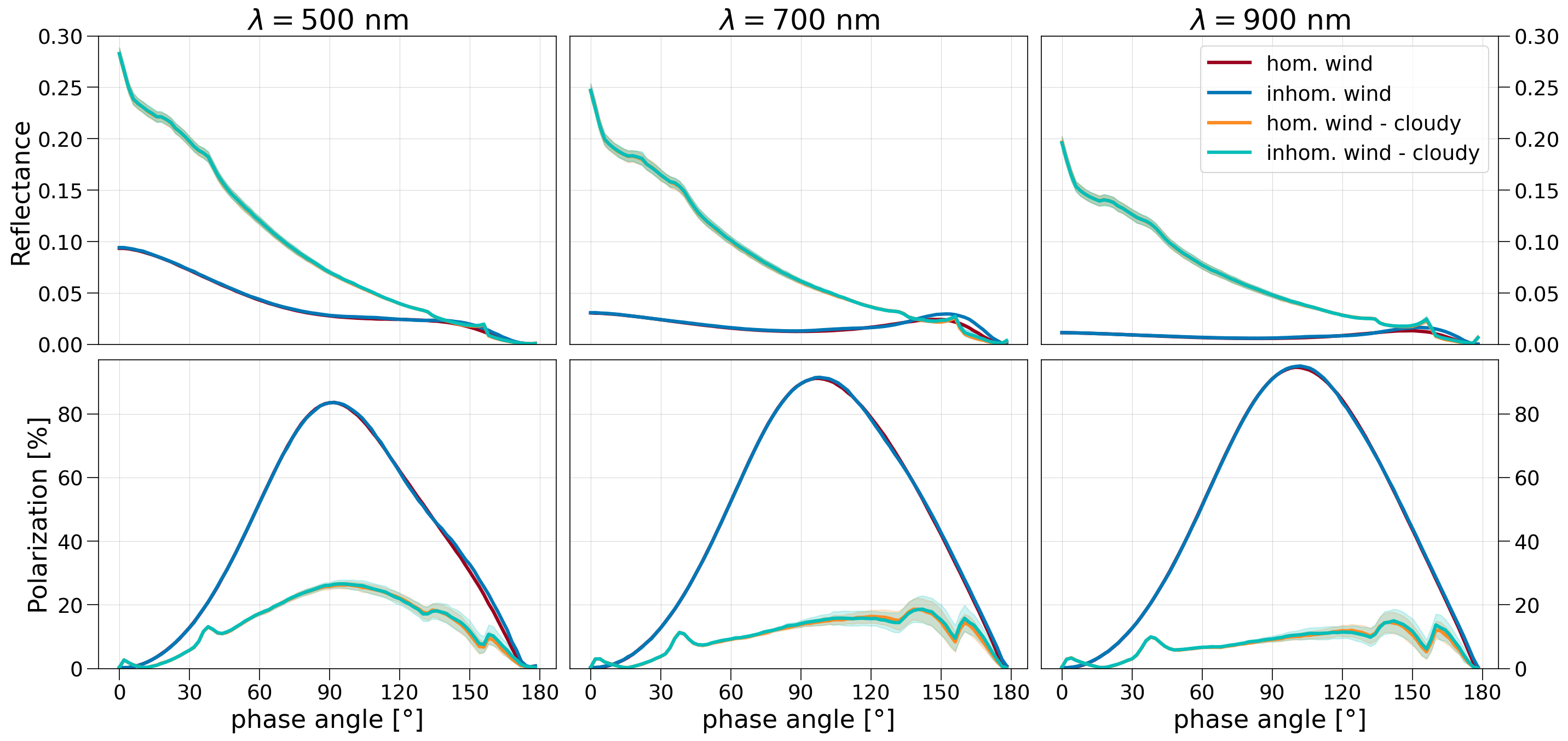}
    \caption{Reflected light (first row) and polarized light (second row) phase curves showing the influence of homogeneous and inhomogeneous wind speed maps. Different columns refer to different wavelengths ($\lambda$): 500, 700, 900~nm.}
    \label{fig:phase_wind}
\end{figure*}

We also discuss the impact of changing the wind speed over the surface for the Ocean planet configuration. Fig. \ref{fig:images_wind} shows the effect of changing the surface wind speed on the ocean glint feature. While the integrated brightness of the ocean glint remains the same, it gets distributed over a larger area when the surface of the ocean gets rougher due to increased wind speed. In the second row of Fig. \ref{fig:images_wind} we show how the ocean glint appears with surface wind speed maps data from the ERA5 reanalysis product. Including realistic wind speed maps, the ocean glint feature does not appear to be symmetric anymore, and shows different features due to weather patterns over the ocean. \\
\noindent Although different wind speeds do not have any impact on the total reflectance of the planet, this might change when simulating cloudy exoplanets, as different spatial distributions of clouds can obscure the ocean glint. In Fig. \ref{fig:spectra_wind} we show the impact of homogeneous and inhomogeneous (from the ERA5 renalysis product) wind speed maps over the reflected and polarized spectra. As expected, for simulations without clouds we did not find any difference among homogeneous and inhomogeneous surface winds, and we find also no significant differences among the 1$\sigma$ cloud spread. A similar behaviour is also observed in the reflected and polarized phase curves (Fig. \ref{fig:phase_wind}), where we find no significant impact due to homogeneous or inhomogeneous wind speed maps.

\section{Impact of albedo seasonal variability}
\label{sec:appendix_E}

\begin{figure*}[h]
    \centering
    \includegraphics[width=1\linewidth]{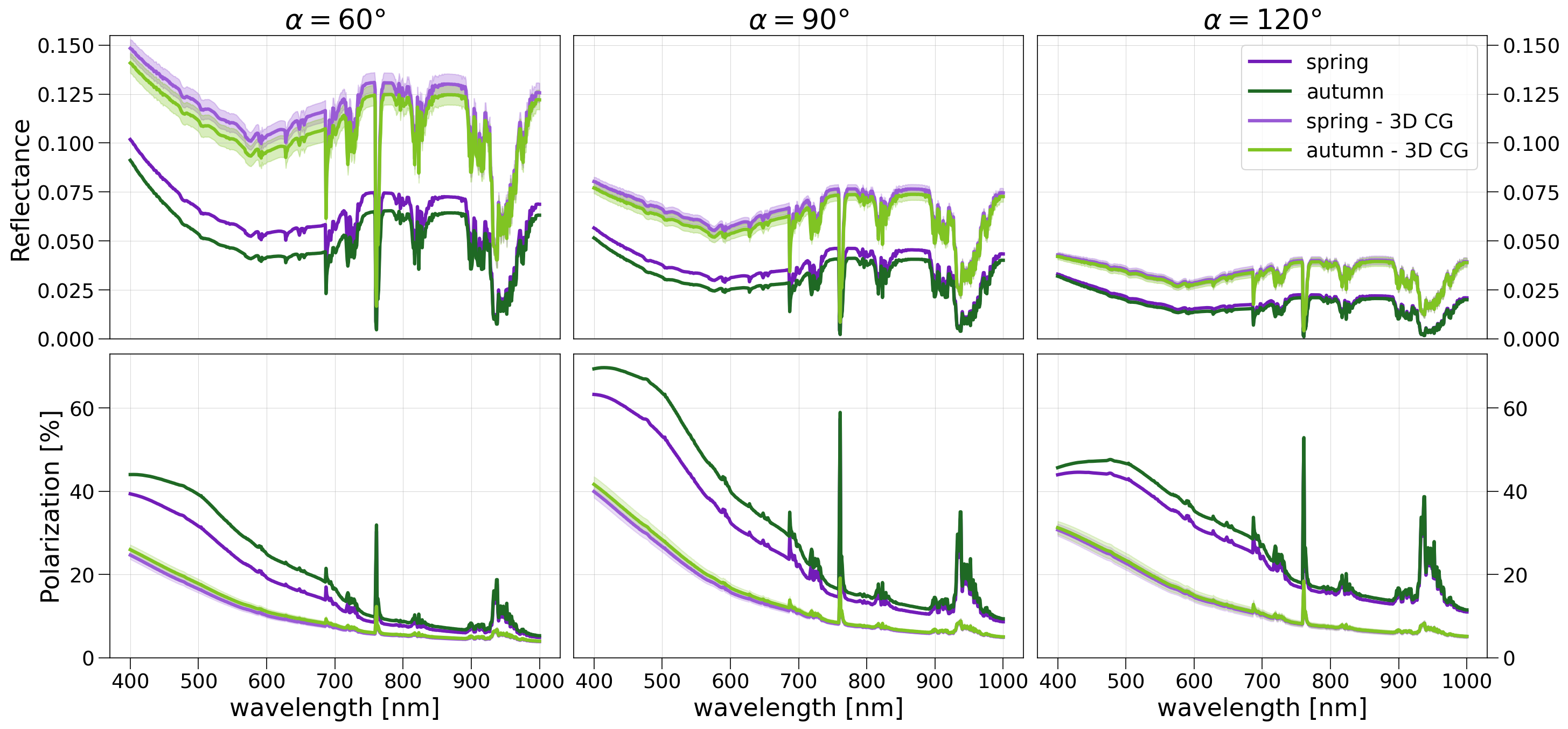}
    \caption{Reflected light (first row) and polarized light (second row) spectra showing the influence of surface albedo seasonal variability. Different columns refer to spectra at different phase angles ($\alpha$): 60, 90, 120\degr.}
    \label{fig:spectra_albedo}
\end{figure*}

\begin{figure*}[h]
    \centering
    \includegraphics[width=1\linewidth]{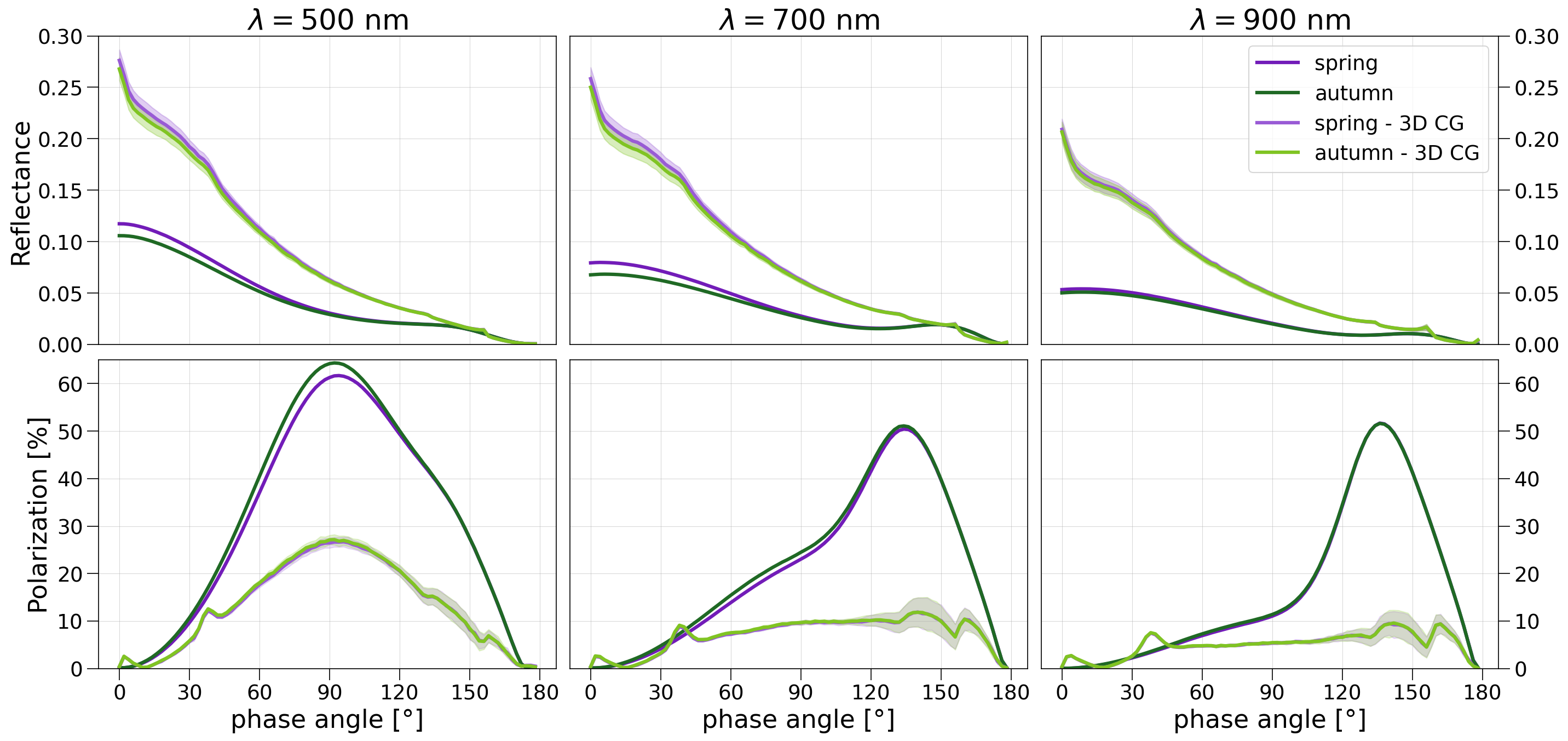}
    \caption{Reflected light (first row) and polarized light (second row) phase curves showing the influence of surface albedo seasonal variability. Different columns refer to different wavelengths ($\lambda$): 500, 700, 900~nm.}
    \label{fig:phase_albedo}
\end{figure*}

Using hyperspectral albedo maps from the HAMSTER dataset \citep{Roccetti2024} for different days of the year (DOYs), we examine the impact of seasonal variability on the planet's surface albedo. We compare between the DOY 080 (spring equinox) and DOY 266 (autumn equinox). As shown in \cite{Roccetti2024}, there is an increase in the overall planet reflectance in the spring due to the fact that the Northern Hemisphere, which hosts almost 80\% of land surface of the planet, exhibits a more significant snow coverage compared to DOY 266, which increases the reflectance of the planet in the visible wavelength range. This is what we also find in Fig. \ref{fig:spectra_albedo} for the reflected light spectra, where in the cloudless scenario we find a larger reflectance for the spring albedo case. The largest spread is found for the $\alpha$ = 60\degr, which also presents the largest illuminated fraction of the planet, thus the largest differences between snow coverage among spring and autumn. When introducing clouds with the 3D CG treatment, we find that the seasonal variability patterns due to surface albedo cannot be distinguished anymore, since they are inside the 1$\sigma$ cloud variability spread for both reflected and polarized light spectra.\\
\noindent In the phase curve comparison (Fig. \ref{fig:phase_albedo}) we find a similar trend than in the reflectance spectra, showing the seasonal variability spread of land surface albedo only in the cloudless simulations. A larger spread is shown at smaller wavelengths, as expected by the typical spectral shape of snow and ice surfaces. In polarization, the spring scenario shows less polarized signal than the autumn case, since polarization behaves as the inverse of the reflectance. In addition, for the cloudless scenario, we also observe a shift in the Rayleigh scattering peak between the spring and autumn cases. However, as for the spectra, all these differences cannot be distinguished anymore adding clouds in the simulations. 

\section{Spectral albedo of the Ocean and Earth-like planet scenarios}
\label{sec:appendix_F}

\begin{figure*}
    \centering
    \includegraphics[width=1\linewidth]{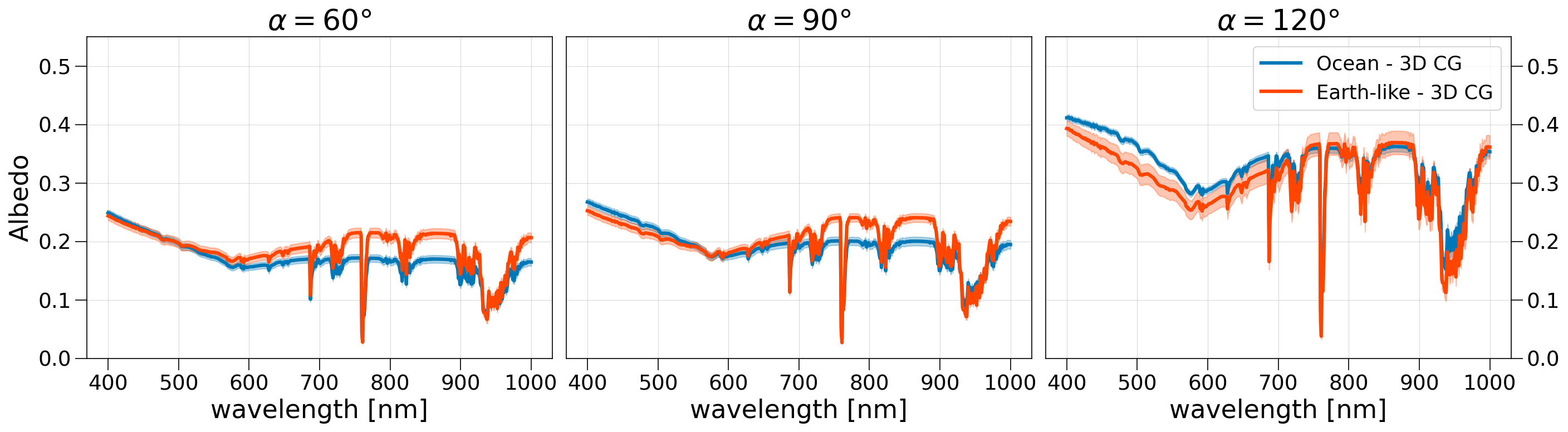}
    \caption{Comparison among the spectral albedo of the Ocean and Earth-like planet scenarios. Different columns refer to different phase angles ($\alpha$): 60, 90, 120\degr. The albedo was obtained from the reflectance by dividing by a labertian phase function (Eq. \ref{eq:lamb_phase})}
    \label{fig:albedo_comparison}
\end{figure*}

In Fig. \ref{fig:albedo_comparison} we show the same spectra as in Fig. \ref{fig:spectra_comparison}, but now as an albedo rather then a reflectance, by simply dividing the reflectance by a lambertian phase function (Eq. \ref{eq:lamb_phase}).

\end{appendix}

\end{document}